\documentclass{amsart}
\usepackage{amsfonts}

\usepackage{graphicx}

\input{tcilatex}
\input tcilatex

\begin{document}
\title[quadratic interactions]{On discrete evolutionary dynamics driven by quadratic interactions}
\author{N. Grosjean, Th. Huillet, G. Rollet}
\address{Laboratoire de Physique Th\'{e}orique et Mod\'{e}lisation\\
CNRS-UMR 8089 et Universit\'{e} de Cergy-Pontoise\\
2 Avenue Adolphe Chauvin, F-95302, Cergy-Pontoise, France\\
E-mail: Nicolas.Grosjean@u-cergy.fr, Thierry.Huillet@u-cergy.fr,
Genevieve.Rollet@u-cergy.fr}
\maketitle

\begin{abstract}
After an introduction to the general topic of models for a given locus of a
diploid population whose quadratic dynamics is determined by a fitness
landscape, we consider more specifically the models that can be treated
using genetic (or train) algebras. In this setup, any quadratic offspring
interaction can produce any type of offspring and after the use of specific
changes of basis, we study the evolution and possible stability of some
examples. We also consider some examples that cannot be treated using the
framework of genetic algebras. Among these are bistochastic matrices.
\end{abstract}


\section{Introduction}

In Section $2$, we briefly revisit the basics of the deterministic dynamics
arising in discrete-time asexual multiallelic evolutionary genetics driven
only by fitness, \emph{in the diploid case with }$K$\emph{\ alleles}. \emph{%
In this setup}, there is a deterministic updating dynamics of the full array
of the genotype frequencies \emph{involving} the fitness matrix attached to
the genotypes. When mating is random so that the Hardy-Weinberg law applies,
one \emph{may limit oneself} to the induced marginal allelic frequencies
dynamics. \emph{Assuming non-overlapping generations,} the updating dynamics
on the simplex \emph{involves the ratio of marginal fitnesses (as affine
functions in the frequencies) and the mean fitness as a quadratic form in
the current frequencies.} We will also consider an alternative updating
mechanism of allelic frequencies over the simplex, namely the Mendelian
segregating mechanism: here the fitness matrix is based on skew-symmetric
matrices and the fitness landscape will be said flat. In the latter flat
fitness model, the offspring can only repeat the genotype of any one of its
parents as is the case in a (fair or unfair) Mendelian inheritance framework.

In Section $3$, we will consider \emph{more }general quadratic interaction
models for which any pair-wise interaction can produce any type of
offspring, thereby generalizing the latter flat fitness model: recombination
is allowed. Under some stochasticity condition on the interactions, the
framework of such models is the one of genetic algebras formalism that we
introduce and develop in some details, \emph{largely inspired by the
fundamental treatises \cite{Wo}, \cite{Lyu}.} In some
(``Gonshor-linearizable'') cases, such dynamics are amenable to linear ones
but in higher dimension. We give five examples for which detailed
computations of the linearization procedure and the precise corresponding
equilibria sets are supplied: the hypergeometric polyploidy model, the
binomial Fisher-Wright model, the Hilbert matrix model, the shift model and
the unbalanced Mendelian model with crossover. The equilibria sets are
shown, depending on the examples, to be either a point, or a curve or a
surface. This concerns Subsection $3.1$.

While using negatively the algebraic criteria that ensures the
Gonshor-linearizability, we give, in Subsection $3.2$, some important
examples where linearizability fails: this includes permutation and more
generally bistochastic models, together with the unbalanced Mendelian
inheritance model (without crossover). The simple $K=2$ dimensional case
will be given a full detailed analysis in this respect.

\section{Single locus: diploid population with $K$ alleles driven by fitness}

For this approach on fitness, we refer to the general treatises \cite{Ew}
and \cite{K1}.

\subsection{Joint and marginal allelic dynamics (fitness)}

Consider $K$ alleles $A_{k}$, $k\in \{1,...,K\}$ attached to a single locus.
Let $W=(W_{k,l}\geq 0:k,l\in \{1,...,K\}^{2})$ where $W_{k,l}$ stands for
the absolute fitness of the genotypes $A_{k}A_{l}$ attached to a single
locus. Since $W_{k,l}$ is proportional to the probability of an $A_{k}A_{l}$
surviving to maturity, it is natural to assume that $W$ is symmetric. Let $%
X=(x_{k,l}:k,l\in \{1,...,K\}^{2})$ be the current frequency distribution at
(integral) time $t$ of the genotypes $A_{k}A_{l}$, so with $x_{k,l}\geq 0$
and $\sum_{k,l}x_{k,l}=1$. Assuming Hardy-Weinberg proportions, the
frequency distribution at time $t$ of the genotypes $A_{k}A_{l}$ is given
by: $x_{k,l}=x_{k}x_{l}$ where $x_{k}=\sum_{l}x_{k,l}$ is the marginal
frequency of allele $A_{k}$. The whole frequency information is now enclosed
within $\mathbf{x}=X\mathbf{1}$ \footnote{%
Throughout, a boldface variable, say $\mathbf{x}$, will represent a
column-vector and its transpose, say $\mathbf{x}^{\prime }$, will be a
row-vector. And $B^{\prime }$ will denote the transpose of some square
matrix $B$. We put $|\mathbf{x}|:=\sum_{k=1}^{K}\left| x_{k}\right| $ and $%
\mathbf{x}\succeq \mathbf{0}$ means that all entries of $\mathbf{x}$ are
nonnegative.}, where $\mathbf{1}^{\prime }=(1,...,1)$ is the $1$-row vector
of dimension $K$. And $\mathbf{x}:=(x_{k}:k\in \left\{ 1,...,K\right\} )$
belongs to the $K-$simplex 
\begin{equation*}
S_{K}=\{\mathbf{x}:=(x_{k}:k=1,...,K)\in \Bbb{R}^{K}:\mathbf{x}\succeq 
\mathbf{0},|\mathbf{x}|=1\}.
\end{equation*}
Define the frequency-dependent marginal fitness of $A_{k}$ by $w_{k}(\mathbf{%
x})=(W\mathbf{x})_{k}:=\sum_{l}W_{k,l}x_{l}$. For some vector ${\mathbf{x}}$%
, denote by $D_{\mathbf{x}}=$diag$\left( x_{k}:k\in \{1,...,K\}\right) $ the
associated diagonal matrix. Assuming non-overlapping generations, the
marginal mapping $\mathbf{p}:S_{K}\to S_{K}$ of the dynamics of $\mathbf{x}$
when driven by viability selection is given by: 
\begin{equation}
\mathbf{x}(t+1)=\mathbf{p}(\mathbf{x}(t))\text{, where }\mathbf{p}({\mathbf{x%
}})=\frac{1}{\omega (\mathbf{x})}D_{\mathbf{x}}W\mathbf{x}=\frac{1}{\omega (%
\mathbf{x})}D_{W\mathbf{x}}\mathbf{x}\,.  \label{15}
\end{equation}
It involves a multiplicative quadratic interaction between $x_{k}$ and $(W%
\mathbf{x})_{k}$, the $k$th entry of the image $W\mathbf{x}$ of $\mathbf{x}$
by $W$ and a normalization by the mean fitness quadratic form $\omega (%
\mathbf{x})=\mathbf{x}^{\prime }W\mathbf{x}$.

\textbf{Recombination.} Genetic recombination is the production of offspring
with combinations of traits that can differ from those found in either
parent. The model (\ref{15}) is a particular case of the following more
general one displaying recombination effects, \cite{Burg}, \cite{Lyu}: let $%
\Gamma _{k}$, $k=1,...,K$ be $K$ nonnegative matrices with entries $\Gamma
_{k}\left( i,j\right) $ representing the propensities for an interacting
pair of alleles of type-$\left( i,j\right) $ to produce a type-$k$ allele.
Let $\Gamma =\sum_{k=1}^{K}\Gamma _{k}$. Consider the dynamics $\mathbf{p}$
on $S_{K}$: 
\begin{equation}
x_{k}(t+1)=p_{k}(\mathbf{x}(t))\text{, where }p_{k}(\mathbf{x})=\frac{%
\mathbf{x}^{^{\prime }}\Gamma _{k}\mathbf{x}}{\mathbf{x}^{^{\prime }}\Gamma 
\mathbf{x}}\text{, }k=1,...,K.  \label{15g}
\end{equation}
In such generalized models, it requires a pair of alleles to produce
offsprings and any pair can in principle produce any type of offspring. The
updating mechanism $\mathbf{p}$ is a fractional transformation with
numerator and denominator both homogeneous of degree two as in (\ref{15}).
Clearly, the mapping $\mathbf{x\rightarrow p}(\mathbf{x})$ is $k$%
-Lipschitzian for $0<k<\infty $, so uniformly continuous on $S_{K}$, so if $%
\mathbf{x}(t)\underset{t\rightarrow \infty }{\rightarrow }\mathbf{x}_{eq}$, $%
\mathbf{x}_{eq}$ has to be a fixed point of $\mathbf{p}$. This fixed point
is unique if $k<1$ but its stability condition is then open. For some very
particular choices of $\Gamma _{k}$, the situation \emph{turns out} to be
simpler. Let for instance $\mathbf{\gamma }_{k}=\Gamma _{k}\mathbf{1}$ and
substitute $P_{k}:=D_{\mathbf{\gamma }_{k}}^{-1}\Gamma _{k}$ to $\Gamma _{k}$
in (\ref{15g}), namely consider the normalized dynamics on $S_{K}$: 
\begin{equation}
x_{k}(t+1)=p_{k}(\mathbf{x}(t))\text{, where }p_{k}(\mathbf{x})=\frac{%
\mathbf{x}^{^{\prime }}P_{k}\mathbf{x}}{\mathbf{x}^{^{\prime }}P\mathbf{x}}%
\text{, }k=1,...,K.  \label{15s}
\end{equation}
Then $P_{k}\mathbf{1}=\mathbf{1}$, $k=1,...,K$, so all $P_{k}$ are
stochastic matrices, not symmetric. And the barycenter $\mathbf{x}%
_{eq}=K^{-1}\mathbf{1}$ is an equiprobable equilibrium state of (\ref{15s}).
Similarly, if $\left\| \Gamma _{k}\right\| _{1}\mathbf{:}=\sum_{i,j}\Gamma
_{k}\left( i,j\right) =$Cte, for all $k=1,...,K$ (all $\Gamma _{k}$ matrices
share the same matrix $1$-norm), then $\mathbf{x}_{eq}=K^{-1}\mathbf{1}$ is
an equilibrium state as well.

Let us now see under what conditions the generalized model (\ref{15g}) boils
down to (\ref{15}). Let $I_{k}$ be the matrix whose entries are all zero
except for the entry in position $\left( k,k\right) $, which is $1$. Suppose 
$\Gamma _{k}=I_{k}W$ where $W$ is the symmetric fitness matrix in (\ref{15}%
). Then $\sum_{k=1}^{K}\Gamma _{k}=\Gamma =W$ is symmetric, $\Gamma _{k}%
\mathbf{x=}\left( W\mathbf{x}\right) _{k}\mathbf{e}_{k}$ where $\mathbf{e}%
_{k}$ is the $k$-th unit vector of $S_{K}$ and (\ref{15g}) matches with (\ref
{15}). If $\Gamma _{k}=I_{k}W$, the propensities for a pair of individuals
of type-$\left( i,j\right) $ to produce a type $k$-individual is zero unless 
$i=k$: a model of Mendelian inheritance. A stochastic version of a similar
model, coined the Fisher-Wright-Haldane model, \emph{was studied in \cite
{Kes1}} and \cite{Kes}. \emph{A general dynamical theory of selection in
multiallelic locus and even of additive selection in multiallelic multilocus
system is developed in \cite{Lyu}, Chapter }$9$\emph{.}

\subsection{The flat fitness model}

We now address the so-called flat fitness model. Let $A$ be \emph{some} real
skew-symmetric matrix, so obeying $A^{\prime }=-A.$ Let $J:=\mathbf{11}%
^{\prime }$ be the all-ones matrix and let $\sigma >0$. Consider the
evolutionary dynamics of the form (\ref{15}) but now when $W$ is of the form 
$W=J+\sigma A\succeq \mathbf{0}$ with $A^{\prime }=-A$ and such that $%
|A_{k,l}|\leq 1/\sigma $. The mean fitness function $\omega (\mathbf{x})$
appearing in (\ref{15}) is a constant $\omega (\mathbf{x})=\mathbf{x}%
^{\prime }W\mathbf{x}=1$, and in this sense the fitness matrix $W$ is called
flat. Because $W_{k,l}+W_{l,k}=2$, these models correspond to constant-sum
games in which each pair of two players has opposed interest or to evolution
under the effect of segregation \emph{distortion} in population genetics;
See \cite{Weis}, \cite{Karlin} and \cite{Hof}. The dynamics (\ref{15}) for
this particular form of $W$ boils down to 
\begin{equation}
\mathbf{x}(t+1)=\mathbf{p}(\mathbf{x}(t))\text{, where }\mathbf{p}({\mathbf{x%
}})=\frac{1}{\omega (\mathbf{x})}D_{\mathbf{x}}W\mathbf{x}=\mathbf{x+}\sigma
D_{\mathbf{x}}A\mathbf{x}.  \label{dyn}
\end{equation}
Let $\Gamma _{k}$, $k=1,...,K$ be $K$ nonnegative symmetric matrices with $%
\left[ 0,1\right] $-valued entries $\Gamma _{k}\left( i,j\right) $
representing the probabilities for a pair of alleles of type-$\left(
i,j\right) $ to produce a type-$k$ allele. Let $\Gamma =\sum_{k=1}^{K}\Gamma
_{k}$ and suppose $\Gamma =J$. Consider the dynamics on $S_{K}$ generalizing
(\ref{dyn}): 
\begin{equation}
x_{k}(t+1)=p_{k}(\mathbf{x}(t))\text{, where }p_{k}(\mathbf{x})=\frac{%
\mathbf{x}^{^{\prime }}\Gamma _{k}\mathbf{x}}{\mathbf{x}^{^{\prime }}\Gamma 
\mathbf{x}}=\mathbf{x}^{^{\prime }}\Gamma _{k}\mathbf{x}\text{, }k=1,...,K.
\label{dyngdv}
\end{equation}
Here $\mathbf{x}^{^{\prime }}\Gamma \mathbf{x}=1$ and the fitness landscape
is flat as in (\ref{dyn}). If in addition $\Gamma _{k}\mathbf{1}=\mathbf{1}$%
, $k=1,...,K$ (all $\Gamma _{k}$ are symmetric bistochastic matrices 
\footnote{%
Symmetric bistochastic matrices is the convex hull of extremal matrices of
the form $\left( P+P^{\prime }\right) /2$ where $P$ is any permutation
matrix.}), or if $\sum_{i,j}\Gamma _{k}\left( i,j\right) =$Cte for all $%
k=1,...,K$, then $\mathbf{x}_{eq}=K^{-1}\cdot \mathbf{1}$ is an unstable
polymorphic equilibrium state of (\ref{dyngdv}), the barycenter of $S_{K}$.

If $\Gamma _{k}\left( i,j\right) =0$ unless $i=k$ or $j=k$ (the offspring
can only repeat the genotype of any one of its parents as in a Mendelian
model), then (\ref{dyngdv}) is of the form (\ref{dyn}) with $A\left(
k,l\right) =2\Gamma _{k}\left( k,l\right) -1$ for $k\neq l$ and $A\left(
l,k\right) =-A\left( k,l\right) $, $\left| A\left( k,l\right) \right| \leq 1$%
, (resulting from $\Gamma _{k}\left( k,l\right) +\Gamma _{l}\left(
l,k\right) =1$), corresponding to a fitness matrix $W=J+\sigma A\succeq 
\mathbf{0}$ with $\sigma =1$. Therefore (\ref{dyn}) is a very particular
case of (\ref{dyngdv}).

\section{Genetic algebras}

In this Section, we will consider the general model (\ref{dyngdv}) under the
flat fitness condition $\mathbf{x}^{^{\prime }}\Gamma \mathbf{x}=1$ which
can be dealt with through genetic algebras ideas, \cite{Wo}, \cite{Lyu}.

Let $\left( \mathbf{e}_{1},...,\mathbf{e}_{K}\right) $ be the natural basis
of $\mathcal{A}=\Bbb{R}^{K}$ representing the extremal states of the simplex 
$S_{K}$. With $\mathbf{x}\left( t\right) \in S_{K}$, we have 
\begin{equation}
\mathbf{x}\left( t\right) =\sum_{k=1}^{K}x_{k}\left( t\right) \mathbf{e}_{k},
\label{B1}
\end{equation}
the species frequency vector in the simplex. Suppose a $K-$dimensional
algebra $\mathcal{A}$ over the field $\Bbb{R}$ with natural multiplication
table 
\begin{equation}
\mathbf{e}_{i}\mathbf{e}_{j}=\sum_{k=1}^{K}\gamma _{ijk}\mathbf{e}_{k},
\label{B2}
\end{equation}
where $\gamma _{ijk}\in \left[ 0,1\right] $ constitute the structure
constants, obeying the property $\sum_{k=1}^{K}\gamma _{ijk}=1$ for all $%
i,j=1,...,K$. $\mathcal{A}$ can be equipped with a weight homomorphism $%
\varpi :$ $\mathcal{A}\rightarrow \Bbb{R}$ obeying $\varpi \left( \mathbf{xy}%
\right) =\varpi \left( \mathbf{x}\right) \varpi \left( \mathbf{y}\right) $
and for which $\forall i$, $\varpi \left( \mathbf{e}_{i}\right) =1$. And
then $S_{K}=\varpi ^{-1}\left( 1\right) \cap \left\{ \mathbf{x}\succeq 
\mathbf{0}\right\} $. Consider the dynamics $\mathbf{x}\left( t+1\right) =%
\mathbf{x}\left( t\right) ^{2}$ (the second-order principal power of $%
\mathbf{x}\left( t\right) $ in the algebra). Identifying $\gamma
_{ijk}=\Gamma _{k}\left( i,j\right) $ and observing $\mathbf{x}^{^{\prime
}}\Gamma \mathbf{x}=1$ as a result of $\Gamma =J$, we obtain (\ref{dyngdv})
evolving in $S_{K}$. Note that, without loss of generality for the dynamics
above, $\gamma _{ijk}=\gamma _{jik}$, a commutativity property ($\mathbf{e}%
_{i}\mathbf{e}_{j}=\mathbf{e}_{j}\mathbf{e}_{i}$). And because in general $%
\left( \mathbf{e}_{i}\mathbf{e}_{j}\right) \mathbf{e}_{k}\neq \mathbf{e}%
_{i}\left( \mathbf{e}_{j}\mathbf{e}_{k}\right) $, $\mathcal{A}$ is
commutative but not associative; such an algebra is called algebra with
genetic realization in \cite{Wo}, \cite{Reed}, or stochastic algebra in \cite
{GMR}. Note also $\mathbf{x}\left( t+m\right) =:\mathbf{x}\left( t\right)
^{\left[ m+1\right] }=\mathbf{x}\left( t\right) ^{2^{m}}$ with $\mathbf{x}%
^{\left[ m\right] }=\mathbf{x}^{\left[ m-1\right] }\mathbf{x}^{\left[
m-1\right] }$, $\mathbf{x}^{\left[ 1\right] }=\mathbf{x},$ defining the
plenary powers of $\mathbf{x}$ in $\mathcal{A}$, not to be confused with the
principal powers of $\mathbf{x}$ in $\mathcal{A}$, namely $\mathbf{x}^{m}=%
\mathbf{xx}^{m-1}$, $\mathbf{x}^{1}=\mathbf{x}$.

Defining $\widehat{\mathbf{e}}_{i}$ to be the multiplication of $\mathbf{x}%
\in \mathcal{A}$ by $\mathbf{e}_{i}$: $\mathbf{x}\overset{\widehat{\mathbf{e}%
}_{i}}{\mapsto }\mathbf{e}_{i}\mathbf{x}$, we get that its corresponding
linear $K\times K$ transformation matrix acting to the left on column
vectors is the matrix $E_{i}$ with entries $E_{i}\left( k,j\right) =\gamma
_{ijk}.$ The matrices $E_{i}$ are all column stochastic ($\forall i,j$, $%
\sum_{k}E_{i}\left( k,j\right) =1$) and they do not commute in general.%
\newline

Let $\left( \mathbf{c}_{1},...,\mathbf{c}_{K}\right) $ denote some canonical
basis in which $\mathbf{x}\left( t\right) =\sum_{k=1}^{K}y_{k}\left(
t\right) \mathbf{c}_{k}.$ Suppose the multiplication table of the $\mathbf{c}%
_{k}$s is given by 
\begin{equation}
\mathbf{c}_{i}\mathbf{c}_{j}=\sum_{k=1}^{K}\lambda _{ijk}\mathbf{c}_{k},
\label{B3}
\end{equation}
where the canonical structure constants $\lambda _{ijk}$ satisfy the Gonshor
conditions \cite{Gon1} 
\begin{equation}
\begin{array}{c}
\lambda _{111}=1 \\ 
\lambda _{1jk}=\lambda _{j1k}=0\text{ if }j>k \\ 
\lambda _{ijk}=0\text{ if }i,j>1\text{and }i\vee j\geq k.
\end{array}
\label{B4}
\end{equation}
If there is a change of basis $\mathbf{e}\rightarrow \mathbf{c}$ so that the
latter Gonshor conditions holds, then $\mathcal{A}$ is called a genetic
algebra. For genetic algebras, it holds that $\varpi \left( \mathbf{c}%
_{1}\right) =1$ and $\varpi \left( \mathbf{c}_{i}\right) =0$, $i=2,...,K$ so
that $I:=\varpi ^{-1}\left( 0\right) =$Ker$\varpi $ is an ideal of $\mathcal{%
A}$ ($I\mathcal{A}\subseteq I$) and $I=$Span$\left( \left\{ \mathbf{c}%
_{2},..,\mathbf{c}_{K}\right\} \right) =:\left\langle \mathbf{c}_{2},..,%
\mathbf{c}_{K}\right\rangle $ is nilpotent ($I^{n}=\left\langle \mathbf{0}%
\right\rangle $ for some integer $n$, the degree of nilpotency). For a
genetic algebra to be a special train algebra, the following additional
condition is required, \cite{Gon1}, \cite{Reed}:

All the principal power subalgebras $I^{m}$ of $\mathcal{A}$ are ideals of $%
\mathcal{A}\Rightarrow \mathcal{A}\supset I\supset ...\supset I^{r}\supset
I^{r+1}=\left\langle \mathbf{0}\right\rangle $ and the sequence of ideals
terminates after $r$ steps called the rank of the special train algebra.

Special train algebras constitute a subclass of train algebras. For train
algebras, the weaker nilpotency condition holds: every element of $I=$Ker$%
\varpi $ is nilpotent of index less or equal $r$. Consequently, if $\mathcal{%
A}$ is a train algebra, for each $\mathbf{x\in }\mathcal{A}$, $r\left( 
\mathbf{x}\right) :=\mathbf{x}\left( \mathbf{x}-\lambda _{1}\right)
...\left( \mathbf{x}-\lambda _{r-1}\right) =\mathbf{0}$ and for each $%
\mathbf{x\in }$Ker$\varpi $, $\mathbf{x}^{r}=\mathbf{0}$; $r\left( \mathbf{x}%
\right) $ is the rank polynomial of $\mathcal{A}$ and the $\lambda _{i}$ are
the principal train roots of $\mathcal{A}$. When $\mathcal{A}$ is moreover a
genetic algebra, the right train roots of $\mathcal{A}$ are $\lambda _{1ii}$%
, $i=1,...,K$ , and the principal train roots of $\mathcal{A}$, as a train
algebra, is a subset of the right train roots of $\mathcal{A}$ (one of which
being $1$), possibly including multiplicities. Apart from $\lambda _{111}=1$%
, all train roots $\lambda _{1ii}$ of a genetic algebra obey $\left| \lambda
_{1ii}\right| \leq 1/2$ (\cite{Wo2}, Coroll. 5). \emph{In this context,} 
\emph{we recall the following general useful result stated in (\cite{Lyu},
theorem }$\emph{7.2.6}$\emph{): ``suppose all the train roots of a genetic
algebra are real. All trajectories converge if and only if all train roots
different from }$\emph{1/2}$\emph{\ lie in the open circle of radius }$\emph{%
1/2}$\emph{\ and the dimension of the manifold of non-zero idempotents is
the same as the number of train roots equal to }$\emph{1/2}$\emph{.''}

All genetic algebras are train algebras but not necessarily special train
algebras, \cite{Gon1}, \cite{Gon2}, \cite{Reed}. For an example of a
(Bernstein) genetic algebra which is not special train and a sufficient
condition for a genetic algebra to be a special train algebra, see Ex. $12$
and Th. $13$ of \cite{FI}. See also the Remark of \cite{ACL}, page $14$.%
\newline

For genetic algebras, we can define the matrices $\Lambda _{k}\left(
i,j\right) =\lambda _{ijk}$, with $\Lambda _{k}$ having zero entries for
those $\left( i,j\right) $ obeying the above constraints. Some of the $%
\lambda _{ijk}$ which are non-zero from the above Gonshor constraints can
occasionally be zero in some examples, thereby defining special classes of
genetic algebras.\newline

Defining $\widehat{\mathbf{c}}_{i}$ to be the left-multiplication of $%
\mathbf{x}\in \mathcal{A}$ by $\mathbf{c}_{i}$: $\mathbf{x}\overset{\widehat{%
\mathbf{c}}_{i}}{\rightarrow }\mathbf{c}_{i}\mathbf{x}$, we get for its left
linear $K\times K$ transformation matrices 
\begin{equation*}
C_{i}=\left[ 
\begin{array}{ccccccc}
0 &  &  &  &  &  &  \\ 
\vdots &  &  &  &  &  &  \\ 
0 &  &  &  &  &  &  \\ 
\underline{\lambda _{i1i}} & 0 &  &  &  &  &  \\ 
\lambda _{i1\left( i+1\right) } & \cdots & \lambda _{ii\left( i+1\right) } & 
0 &  &  &  \\ 
\vdots &  & \vdots & \ddots &  &  &  \\ 
\vdots &  & \vdots &  & \ddots &  &  \\ 
\lambda _{i1K} & \cdots & \lambda _{iiK} & \cdots & \cdots & \lambda
_{i\left( K-1\right) K} & 0
\end{array}
\right] \text{ if }i=2,...,K
\end{equation*}
\begin{equation*}
C_{1}=\left[ 
\begin{array}{cccccc}
\underline{\lambda _{111}} &  &  &  &  &  \\ 
\lambda _{112} & \underline{\lambda _{122}} &  &  &  &  \\ 
\vdots &  & \ddots &  &  &  \\ 
\lambda _{11i} & \cdots & \cdots & \underline{\lambda _{1ii}} &  &  \\ 
\vdots &  &  & \vdots & \ddots &  \\ 
\lambda _{11K} & \cdots & \cdots & \lambda _{1iK} & \cdots & \underline{%
\lambda _{1KK}}
\end{array}
\right] \text{ if }i=1.
\end{equation*}
The right train roots $\lambda _{1ii}$ of $\mathcal{A}$ are read on the
diagonal of $C_{1}$ (they are the characteristic roots of the operator which
is multiplication by $\mathbf{c}_{1}$), whereas the left train roots $%
\lambda _{i1i}$ of $\mathcal{A}$ are read on the $\left( i,1\right) -$entry
of $C_{i}$. They are the values which were underlined.

We note that with $\left\{ \mathbf{\omega }_{i,k}\text{, }i=2,...,K\text{, }%
k>i\right\} $ the column $K-$vectors with entries $\mathbf{\omega }%
_{i,k}\left( j\right) =\lambda _{ijk}$, $j=1,...,k-1$, $=0$ if $j=k,...,K$,
so that $\mathbf{\omega }_{i,k}^{\prime }\mathbf{e}_{l}=0$ for all $%
l=k,...,K $, then 
\begin{equation*}
C_{i}=\lambda _{i1i}\mathbf{e}_{i}\mathbf{e}_{1}^{\prime }+\sum_{k=i+1}^{K}%
\mathbf{e}_{k}\mathbf{\omega }_{i,k}^{\prime }\text{, }i=2,...,K.
\end{equation*}
This decomposition into projectors together with the property $\mathbf{%
\omega }_{i,k}^{\prime }\mathbf{e}_{l}=0$ is enough to ensure the nilpotency
of the latter matrices $C_{i}$ and it gives their orders of nilpotency.

From the shape of the $C_{i}$s, it also holds that $\forall i=2,...,K:$ $%
C_{i}\left\langle \mathbf{c}_{K}\right\rangle =\left\langle \mathbf{0}%
\right\rangle $ (all $C_{i}$, $i=2,...,K$ share $\mathbf{c}_{K}$ as a common
eigenvector associated to the eigenvalue $0$) and, with $\left\langle 
\mathbf{c}_{k+1},...,\mathbf{c}_{K}\right\rangle \subset \left\langle 
\mathbf{c}_{k},...,\mathbf{c}_{K}\right\rangle $, $k=1,...,K-1$, 
\begin{eqnarray*}
C_{i}\left\langle \mathbf{c}_{k},...,\mathbf{c}_{K}\right\rangle &\subseteq
&\left\langle \mathbf{c}_{i+1},...,\mathbf{c}_{K}\right\rangle \text{, for
all }i=2,...,K\text{ and }k=2,...i \\
C_{i}\left\langle \mathbf{c}_{k},...,\mathbf{c}_{K}\right\rangle &\subseteq
&\left\langle \mathbf{c}_{k+1},...,\mathbf{c}_{K}\right\rangle \text{, for
all }i=2,...,K\text{ and }k=i,...,K \\
C_{1}\left\langle \mathbf{c}_{k},...,\mathbf{c}_{K}\right\rangle &\subseteq
&\left\langle \mathbf{c}_{k},...,\mathbf{c}_{K}\right\rangle \text{ if }i=1%
\text{ and }k=1,...,K.
\end{eqnarray*}
If $\mathbf{x=}\sum_{j}y_{j}\mathbf{c}_{j}$, where the $y_{j}$s are the
coordinates of $\mathbf{x}\in S_{K}$ in the canonical basis (with $y_{1}=1$%
), the matrix associated to the left multiplication $\widehat{\mathbf{x}}$
by $\mathbf{x}$ is $C_{\mathbf{x}}=\sum_{j}y_{j}C_{j}$, which is lower-left
triangular with diag$\left( C_{\mathbf{x}}\right) =$diag$\left( \lambda
_{1ii}\right) $. Therefore, 
\begin{equation*}
\sum_{j=1}^{K}y_{j}C_{j}\mathbf{x=}\sum_{j,k=1}^{K}y_{j}y_{k}C_{j}\mathbf{c}%
_{k}
\end{equation*}
are the coordinates of $\mathbf{x}^{2}\in S_{K}$ in the canonical basis.%
\newline

Suppose $\mathbf{c}_{i}=\sum_{j=1}^{K}B\left( i,j\right) \mathbf{e}_{j}$ so
with (non-singular) matrix $B$ defining the change of basis. Then $\mathbf{e}%
_{i}=\sum_{j=1}^{K}B^{-1}\left( i,j\right) \mathbf{c}_{j}=\mathbf{c}%
_{1}+\sum_{j=2}^{K}B^{-1}\left( i,j\right) \mathbf{c}_{j}$ with $%
B^{-1}\left( i,1\right) =1$ so as to ensure the compatibility of $\forall i$%
, $\varpi \left( \mathbf{e}_{i}\right) =1$ and $\varpi \left( \mathbf{c}%
_{i}\right) =\delta _{i,1}.$

In the sequel, we shall use 
\begin{equation*}
B_{1}=\left[ 
\begin{array}{llll}
1 &  &  &  \\ 
-1 & 1 &  &  \\ 
\vdots & \mathbf{0} & \ddots &  \\ 
-1 & \mathbf{0} & \mathbf{0} & 1
\end{array}
\right] \text{ with }B_{1}^{-1}=\left[ 
\begin{array}{llll}
1 &  &  &  \\ 
1 & 1 &  &  \\ 
\vdots & \mathbf{0} & \ddots &  \\ 
1 & \mathbf{0} & \mathbf{0} & 1
\end{array}
\right] ,
\end{equation*}
and $B_{2}\left( i,j\right) =\left( -1\right) ^{j-1}\binom{i-1}{j-1},$ with $%
B_{2}^{-1}\left( i,j\right) =B_{2}\left( i,j\right) .$ In the latter case,
we shall also use $B_{3}=B_{2}P$ where $P$ is the permutation matrix $%
P\left( i,j\right) =\delta _{i,K+1-i}$ so with $B_{3}\left( i,j\right)
=\left( -1\right) ^{K-j}\binom{i-1}{K-j}.$

Write $b_{ij}:=B\left( i,j\right) .$ Then (using Einstein notations while
summing over repeated indices): $\lambda _{ijk}=b_{ii^{\prime
}}b_{jj^{\prime }}\gamma _{i^{\prime }j^{\prime }k^{\prime }}b_{k^{\prime
}k}^{-1}$ gives the way the natural structure constants are deformed into
the canonical ones of Gonshor, with the obvious inverse transformation,
would the algebra be genetic. Note that this also means $C_{i}=B^{\prime
-1}\left( \sum_{i^{\prime }}b_{ii^{\prime }}E_{i^{\prime }}\right) B^{\prime
}$ where $B^{\prime }$ is the transpose of $B$, together with 
\begin{equation}
E_{i}=B^{\prime }\left( \sum_{i^{\prime }}b_{ii^{\prime }}^{-1}C_{i^{\prime
}}\right) B^{^{\prime }-1}=B^{\prime }\left( C_{1}+\sum_{i^{\prime }\neq
1}b_{ii^{\prime }}^{-1}C_{i^{\prime }}\right) B^{^{\prime }-1}.  \label{B5}
\end{equation}
The latter identity shows that for genetic algebras, the $E_{i}$s must be
mutually similar to triangular matrices (non-commutative in general and
simultaneously triangularizable by the same similarity matrix $B^{\prime }$%
). Because $\forall i$, $b_{i1}^{-1}=1$, for every $i,j$, 
\begin{equation}
E_{i}-E_{j}=B^{\prime }\left( \sum_{i^{\prime }\neq 1}\left( b_{ii^{\prime
}}^{-1}-b_{ji^{\prime }}^{-1}\right) C_{i^{\prime }}\right) B^{^{\prime }-1},
\label{B6}
\end{equation}
with the matrix inside the parenthesis strictly lower-triangular. Thus $%
E_{i}-E_{j}$ must also be similar to a nilpotent matrix, so nilpotent itself.%
\newline

Given $\lambda _{ijk}$ and $b_{ij}$ it is not always satisfied that $\gamma
_{ijk}$ are $\left[ 0,1\right] -$valued with the property $\sum_{k}\gamma
_{ijk}=1$ for all $i,j$. With $\mathbf{\Gamma :}=\left( \Gamma
_{k},k=1,...,K\right) $, $\mathbf{\Lambda :}=\left( \Lambda
_{k},k=1,...,K\right) $ and $B$, we shall say that the triple $\left( 
\mathbf{\Gamma },\mathbf{\Lambda },B\right) $ is Gonshor-compatible if the $%
\Gamma _{k}$ are $\left[ 0,1\right] $-valued matrices with $\sum_{k}\Gamma
_{k}=J$. In this case, the model $\mathbf{\Gamma }$ is linearizable in a
higher dimensional state-space whose rapidly growing dimension is given in
Proposition $2$ of Abraham \cite{Ab1} (would there be no other zero $\lambda
_{ijk}$ but the ones given from the Gonshor constraints, the dimension of
the embedding linear space grows like $\sqrt{2}^{K^{2}}$).

\subsection{Examples of models akin to a genetic algebra}

Let us give some examples of genetic algebras. \emph{In case the genetic
algebras under study are train algebras, our examples will serve as an
illustration of (\cite{Lyu}, theorem 7.2.6) discussed above, characterizing
the dimension of the equilibria sets.}

$\bullet $ \textbf{Pascal change of basis}: Suppose the hypergeometric model 
$\mathbf{\Gamma }$ with 
\begin{equation}
\gamma _{ijk}=\binom{2\left( K-1\right) }{K-1}^{-1}\binom{i+j-2}{k-1}\binom{%
2K-\left( i+j\right) }{K-k}\text{, }i,j,k=1,...,K.  \label{B7}
\end{equation}
$\gamma _{ijk}$ (as the probability that an $i,j$ interaction produces $k$)
is the probability that $k-1$ successes occur in a $K-1$ draw without
replacement from a population of size $2\left( K-1\right) $ containing $%
i+j-2 $ successes and $2K-\left( i+j\right) $ failures, $2\leq i+j\leq 2K$.
Clearly, $\gamma _{ijk}$ are $\left[ 0,1\right] -$valued as probabilities
with $\sum_{k}\gamma _{ijk}=1$ as a result of the Vandermonde convolution
identity. Then using the change of basis $B_{3}\left( i,j\right) =\left(
-1\right) ^{K-j}\binom{i-1}{K-j}$, we get the Gonshor-like structure
constants 
\begin{equation}
\begin{array}{c}
\lambda _{ijk}=\binom{2\left( K-1\right) }{i+j-2}^{-1}\binom{K-1}{i+j-2}%
\text{, if }k=i+j-1\text{,} \\ 
=0\text{ if not}
\end{array}
\label{B8}
\end{equation}
and using $B_{2}\left( i,j\right) =\left( -1\right) ^{j-1}\binom{i-1}{j-1}$,
with $S_{ijk}:=\sum_{l=0}^{i+j-2}\left( -1\right) ^{l}\binom{i+j-2}{l}\binom{%
l}{k-1}$%
\begin{equation*}
\lambda _{ijk}=\binom{2\left( K-1\right) }{K-1}^{-1}\binom{2K-k-1}{K-k}%
\left( -1\right) ^{k-1}S_{ijk}\text{, }i+j\leq k+1.
\end{equation*}
which are Gonshor-like structure constants. More precisely, because here $%
S_{ijk}=\left( -1\right) ^{k-1}$ if $i+j=k+1$, $=0$ if $i+j\neq k+1$ 
\begin{eqnarray*}
\lambda _{ijk} &=&\binom{2\left( K-1\right) }{K-1}^{-1}\binom{2K-\left(
i+j\right) }{K-\left( i+j-1\right) }\text{, if }i+j=k+1 \\
&=&0\text{, if }i+j\neq k+1.
\end{eqnarray*}

For the hypergeometric model $\mathbf{\Gamma }$, $\left( \mathbf{\Gamma },%
\mathbf{\Lambda },B\right) $ is Gonshor-compatible for $B=B_{2}$ and $%
B=B_{3} $. The latter models are models of polyploidy of degree $1.$ In the
polyploidy of degree $1$ examples, the $\Lambda _{k}$s are zero except on
the anti-diagonals $i+j=k+1$. The genetic polyploidy algebra is a special
train algebra \footnote{%
It can indeed be checked here that $I^{2}=\left\langle \mathbf{c}_{3},...,%
\mathbf{c}_{K}\right\rangle $, $I^{3}=\left\langle \mathbf{c}_{4},...,%
\mathbf{c}_{K}\right\rangle $,...and $\mathcal{A}I^{2}\subseteq I^{2},$ $%
\mathcal{A}I^{3}\subseteq I^{3}$,...} with train roots $\lambda _{1ii}=%
\binom{2\left( K-1\right) }{K-1}^{-1}\binom{2\left( K-1\right) -\left(
i-1\right) }{K-i}$ verifying $\lambda _{111}=1,$ $\lambda _{122}=1/2,$ $%
\lambda _{1\left( i+1\right) \left( i+1\right) }<\lambda _{1ii},$ \cite{Gon1}%
. Because $\lambda _{122}=1/2$ is a train root with multiplicity $1$, we
expect an equilibrium curve, (\cite{Gon1}, \cite{Lyu}\emph{, theorem 7.2.6}).%
\newline

Building from this example the column stochastic matrices $E_{i}$ with
entries $E_{i}\left( k,j\right) =\gamma _{ijk}$, they can be seen to be
simultaneously triangularizable and the matrices $E_{i}-E_{j}$ are all
nilpotent.\newline

\textbf{Example:} Let $K=4$ and consider the Gonshor multiplication table in
this low-dimensional case ($\lambda _{111}=1$) using $B_{3}$. We have 
\begin{eqnarray*}
\mathbf{c}_{1}^{2} &=&\lambda _{111}\mathbf{c}_{1}=\mathbf{c}_{1} \\
\mathbf{c}_{1}\mathbf{c}_{2} &=&\lambda _{122}\mathbf{c}_{2}\text{, }\mathbf{%
c}_{1}\mathbf{c}_{3}=\lambda _{133}\mathbf{c}_{3} \\
\mathbf{c}_{1}\mathbf{c}_{4} &=&\lambda _{144}\mathbf{c}_{4}\text{, }\mathbf{%
c}_{2}^{2}=\lambda _{223}\mathbf{c}_{3}\text{, }\mathbf{c}_{2}\mathbf{c}%
_{3}=\lambda _{234}\mathbf{c}_{4} \\
\mathbf{c}_{2}\mathbf{c}_{4} &=&\mathbf{c}_{3}^{2}=\mathbf{c}_{3}\mathbf{c}%
_{4}=\mathbf{c}_{4}^{2}=0
\end{eqnarray*}
where $\lambda _{ij\left( i+j-1\right) }=\binom{6}{i+j-2}^{-1}\binom{3}{i+j-2%
}$ and so $\lambda _{122}=1/2$, $\lambda _{133}=1/5$, $\lambda _{144}=1/20$, 
$\lambda _{223}=1/5$ and $\lambda _{234}=1/20$. Considering the time
evolution $\mathbf{x}\left( t+1\right) =\mathbf{x}\left( t\right) ^{2}$ in
the Gonshor basis where $\mathbf{x}\left( t\right) =:\mathbf{c}%
_{1}+y_{2}\left( t\right) \mathbf{c}_{2}+y_{3}\left( t\right) \mathbf{c}%
_{3}+y_{4}\left( t\right) \mathbf{c}_{4}$, we get 
\begin{eqnarray*}
\mathbf{x}\left( t+1\right) &=&\mathbf{c}_{1}^{2}+y_{2}^{2}\left( t\right) 
\mathbf{c}_{2}^{2}+2y_{2}\left( t\right) \mathbf{c}_{1}\mathbf{c}%
_{2}+2y_{3}\left( t\right) \mathbf{c}_{1}\mathbf{c}_{3}+2y_{4}\left(
t\right) \mathbf{c}_{1}\mathbf{c}_{4}+2y_{2}\left( t\right) y_{3}\left(
t\right) \mathbf{c}_{2}\mathbf{c}_{3} \\
&=&\mathbf{c}_{1}+2y_{2}\left( t\right) \lambda _{122}\mathbf{c}_{2}+\left(
y_{2}^{2}\left( t\right) \lambda _{223}+2y_{3}\left( t\right) \lambda
_{133}\right) \mathbf{c}_{3}+2\left( y_{4}\left( t\right) \lambda
_{144}+y_{2}\left( t\right) y_{3}\left( t\right) \lambda _{234}\right) 
\mathbf{c}_{4} \\
&=&:y_{1}\left( t+1\right) \mathbf{c}_{1}+y_{2}\left( t+1\right) \mathbf{c}%
_{2}+y_{3}\left( t+1\right) \mathbf{c}_{3}+y_{4}\left( t+1\right) \mathbf{c}%
_{4}.
\end{eqnarray*}
To get a finite recursion, we need to generate the evolution of the
additional states $y_{2}^{2}\left( t\right) $, $y_{2}\left( t\right)
y_{3}\left( t\right) $ and $y_{2}^{3}\left( t\right) $ one of which is
cubic. We get 
\begin{eqnarray*}
y_{2}^{2}\left( t+1\right) &=&4y_{2}^{2}\left( t\right) \lambda _{122}^{2} \\
y_{2}\left( t+1\right) y_{3}\left( t+1\right) &=&2y_{2}^{3}\left( t\right)
\lambda _{122}\lambda _{223}+4y_{2}\left( t\right) y_{3}\left( t\right)
\lambda _{122}\lambda _{133} \\
y_{2}^{3}\left( t+1\right) &=&8y_{2}^{3}\left( t\right) \lambda _{122}^{3}
\end{eqnarray*}
There are three additional states to generate here and we obtain the closed
7-dimensional (triangular) evolution 
\begin{equation*}
\left[ 
\begin{array}{c}
y_{1}\left( t+1\right) \\ 
y_{2}\left( t+1\right) \\ 
y_{2}^{2}\left( t+1\right) \\ 
y_{2}^{3}\left( t+1\right) \\ 
y_{2}y_{3}\left( t+1\right) \\ 
y_{3}\left( t+1\right) \\ 
y_{4}\left( t+1\right)
\end{array}
\right] =\left[ 
\begin{array}{ccccccc}
1 &  &  &  &  &  &  \\ 
0 & 2\lambda _{122} &  &  &  &  &  \\ 
0 & 0 & 4\lambda _{122}^{2} &  &  &  &  \\ 
0 & 0 & 0 & 8\lambda _{122}^{3} &  &  &  \\ 
0 & 0 & 0 & 2\lambda _{122}\lambda _{223} & 4\lambda _{122}\lambda _{133} & 
&  \\ 
0 & 0 & \lambda _{223} & 0 & 0 & 2\lambda _{133} &  \\ 
0 & 0 & 0 & 0 & 2\lambda _{234} & 0 & 2\lambda _{144}
\end{array}
\right] \left[ 
\begin{array}{c}
y_{1}\left( t\right) \\ 
y_{2}\left( t\right) \\ 
y_{2}^{2}\left( t\right) \\ 
y_{2}^{3}\left( t\right) \\ 
y_{2}y_{3}\left( t\right) \\ 
y_{3}\left( t\right) \\ 
y_{4}\left( t\right)
\end{array}
\right] .
\end{equation*}
The transition matrix of this dynamics has $1$ as a dominant eigenvalue with
multiplicity 4 (because $\lambda _{122}=1/2$), the corresponding eigenvector
being, up to an indeterminate constant $y_{2}:$ $\left(
1,y_{2},y_{2}^{2},y_{2}^{3},y_{2}^{3}/3,y_{2}^{2}/3,y_{2}^{3}/27\right) .$

Recalling the correspondence between the $\mathbf{x}$s and the $\mathbf{y}$%
s, namely $x_{k}=\sum_{j=1}^{4}y_{j}B_{3}\left( j,k\right) =\left( -1\right)
^{k}\sum_{j=5-k}^{4}y_{j}\binom{j-1}{4-k}$, in view of $\mathbf{y}%
_{*}^{\prime }=\left( 1,y_{2},y_{2}^{2}/3,y_{2}^{3}/27\right) $, leads to
equilibrium states $\mathbf{x}_{eq}$ of the $\mathbf{x}$s dynamics in the
simplex given by 
\begin{equation*}
\mathbf{x}_{eq}^{\prime }=\left(
-y_{2}^{3}/27;y_{2}^{2}/3+y_{2}^{3}/9;-\left(
y_{2}+2y_{2}^{2}/3+y_{2}^{3}/9\right)
;1+y_{2}+y_{2}^{2}/3+y_{2}^{3}/27\right) ,
\end{equation*}
with normalizing constant $1$ and for those values of $-3\leq y_{2}\leq 0$
for which $\mathbf{x}_{eq}$ belongs to the simplex. This equilibrium curve,
parameterized by $y_{2}$, is cubic and skew; it is stable and the rate at
which the dynamics moves to $\left\{ \mathbf{x}_{eq}\right\} $ is geometric
with parameter $2\left( \lambda _{133}\vee \lambda _{144}\right) =2/5<1$.
Note $\mathbf{x}_{eq}^{\prime }=\left( 1;0;0;0\right) $ if $y_{2}=-3$ and $%
\mathbf{x}_{eq}^{\prime }=\left( 0;0;0;1\right) $ if $y_{2}=0$; they are the
extreme points of the cubics on the simplex.\newline

\textbf{Polyploidy of degree }$d$\textbf{:} let $d\geq 2$ be some integer,
with $2d$ measuring the degree of polyploidy (the case $d=1$ being the
previous case). Suppose the extended hypergeometric model $\mathbf{\Gamma }$
with 
\begin{equation}
\gamma _{ijk}=\binom{2d\left( K-1\right) }{K-1}^{-1}\binom{d\left(
i+j-2\right) }{k-1}\binom{d\left( 2K-\left( i+j\right) \right) }{K-k}\text{, 
}i,j,k=1,...,K.  \label{B9}
\end{equation}
$\gamma _{ijk}$ is the probability that $k-1$ successes occur in a $K-1$
draw without replacement from a population of size $2d\left( K-1\right) $
containing $d\left( i+j-2\right) $ successes and $d\left( 2K-\left(
i+j\right) \right) $ failures, $2\leq i+j\leq 2K$. Then, using the change of
basis $B_{3}$, with $S_{ijk}\left( d\right) :=\sum_{l=0}^{i+j-2}\left(
-1\right) ^{l}\binom{i+j-2}{l}\binom{dl}{k-1}$, we get the Gonshor-like
structure constants 
\begin{equation}
\begin{array}{c}
\lambda _{ijk}=\binom{2d\left( K-1\right) }{i+j-2}^{-1}\binom{d\left(
K-1\right) }{i+j-2}\left( -1\right) ^{k-1}S_{ijk}\left( d\right) \text{, if }%
k\geq i+j-1\text{,} \\ 
=0\text{ if not}
\end{array}
\label{B10}
\end{equation}
and using the change of basis $B_{2}$%
\begin{equation*}
\lambda _{ijk}=\binom{2d\left( K-1\right) }{K-1}^{-1}\binom{2d\left(
K-1\right) -\left( k-1\right) }{K-k}\left( -1\right) ^{k-1}S_{ijk}\left(
d\right) \text{, }i+j\leq k+1.
\end{equation*}
In both cases, $S_{ijk}\left( d\right) $ is such that $S_{ijk}\left(
d\right) \neq 0$ if $i+j\leq k+1$, $=0$ if $i+j>k+1.$ Although more complex,
this is also a Gonshor-like set of structure constants. In particular, in
the latter $B_{2}$ case, when $i+j-1$ varies from $1$ to $K$, in view of ($%
k=i+j-1$) $S_{ijk}\left( d\right) =\left( -d\right) ^{i+j-2}$%
\begin{equation*}
\lambda _{ij\left( i+j-1\right) }=\binom{2d\left( K-1\right) }{K-1}^{-1}%
\binom{2d\left( K-1\right) -\left( i+j-2\right) }{K-\left( i+j-1\right) }%
d^{i+j-2},
\end{equation*}
defining the train roots (right and left train roots being respectively 
\begin{equation*}
\lambda _{1jj}=\binom{2d\left( K-1\right) }{K-1}^{-1}\binom{2d\left(
K-1\right) -\left( j-1\right) }{K-j}d^{j-1}\text{ and }\lambda
_{i1i}=\lambda _{1ii},
\end{equation*}
with $\lambda _{111}=1$, $\lambda _{122}=1/2$, $\lambda _{1\left( i+1\right)
\left( i+1\right) }<\lambda _{1ii}$). In the polyploidy of degree\textbf{\ }$%
d>1$ examples, the $\Lambda _{k}$ are upper-left triangular (a special class
of genetic algebras known as special train genetic algebra with train roots $%
\lambda _{ij\left( i+j-1\right) }$)\textbf{. }Like in the polyploidy model%
\textbf{\ }of degree $d=1$, in both $B_{3}$ and $B_{2}$ cases, the
equilibrium set is a curve because $\lambda _{122}=1/2$ is a train root with
multiplicity $1$, (\cite{Gon2}, \cite{Lyu}\emph{, theorem 7.2.6}.)\newline

\textbf{Fisher-Wright model.} Let $\alpha >0,1>\beta >0$ obeying $\alpha
<2\left( K-1\right) \left( 1-\beta \right) $. Suppose the Fisher-Wright
model $\mathbf{\Gamma }$ for which $i,j,k=1,...,K$ and 
\begin{equation}
\gamma _{ijk}=\frac{\binom{K-1}{k-1}}{\left( 2\left( K-1\right) \right)
^{\left( K-1\right) }}\left( \alpha +\beta \left( i+j-2\right) \right)
^{k-1}\left( 2\left( K-1\right) -\alpha -\beta \left( i+j-2\right) \right)
^{K-k}.  \label{B11}
\end{equation}
$\gamma _{ijk}$ is a binomial-like probability system obeying $%
\sum_{k=1}^{K}\gamma _{ijk}=1$. Using the change of basis $B_{2}$, whenever $%
i+j\leq k+1$, we easily get 
\begin{equation}
\lambda _{ijk}=\left( -1\right) ^{k-1}\frac{\binom{K-1}{k-1}}{\left( 2\left(
K-1\right) \right) ^{\left( k-1\right) }}\sum_{l=0}^{i+j-2}\left( -1\right)
^{l}\binom{i+j-2}{l}\left( \alpha +\beta l\right) ^{k-1}\text{,}  \label{B12}
\end{equation}
which are Gonshor-like structure constants with $\lambda _{ij1}=0$ ($ij\neq
1 $) and $\lambda _{ijk}=0$ if $i+j>k+1$, $\lambda _{ijk}$ depending only on 
$i+j$.

The last point can be checked while observing $\sum_{l=0}^{n}\left(
-1\right) ^{l}\binom{n}{l}l^{k}=0$ for all $0\leq k\leq n-1$: consider
indeed the degree-$n$ polynomial $P_{n}\left( x\right) =\left( x-1\right)
^{n}$ and with $D_{k}=\left( x\partial _{x}\right) ^{k}$ consider then the
degree-$n$ polynomial $D_{k}P_{n}\left( x\right) $. We have $%
D_{k}P_{n}\left( 1\right) =\sum_{l=0}^{n}\left( -1\right) ^{l}\binom{n}{l}%
l^{k}=0$, for all $0\leq k\leq n-1$ and $D_{n}P_{n}\left( 1\right)
=\sum_{l=0}^{n}\left( -1\right) ^{l}\binom{n}{l}l^{n}=n!$. Note that with $%
i+j\leq k+1$, 
\begin{equation*}
\lambda _{ijk}=\binom{K-1}{k-1}\left( \frac{-\alpha }{2\left( K-1\right) }%
\right) ^{k-1}\sum_{l=i+j-2}^{k-1}\binom{k-1}{l}\left( \beta /\alpha \right)
^{l}D_{l}P_{i+j-2}\left( 1\right) .
\end{equation*}
For the Fisher-Wright model $\mathbf{\Gamma }$, $\left( \mathbf{\Gamma },%
\mathbf{\Lambda },B_{2}\right) $ is Gonshor-compatible and this model
defines a special train algebra with right (and left) train roots (when $%
\beta <1$) 
\begin{equation*}
\lambda _{1jj}=\frac{\left( -1\right) ^{j-1}}{\left( 2\left( K-1\right)
\right) ^{\left( j-1\right) }}\binom{K-1}{j-1}\sum_{l=0}^{j-1}\left(
-1\right) ^{l}\binom{j-1}{l}\left( \alpha +\beta l\right) ^{j-1}=\frac{%
\binom{K-1}{j-1}\left( j-1\right) !\beta ^{j-1}}{\left( 2\left( K-1\right)
\right) ^{\left( j-1\right) }},
\end{equation*}
obeying $\lambda _{1\left( j+1\right) \left( j+1\right) }/\lambda
_{1jj}=\beta \left( K-j\right) /\left( 2\left( K-1\right) \right) <1$. In
particular, $\lambda _{122}=\beta /2<1/2,$ $\lambda _{133}=\left( K-2\right)
\left( \beta /2\right) ^{2}/\left( K-1\right) <\left( \beta /2\right)
^{2}<1/4$,...\newline

\textbf{Example:} Let $K=3$ and consider the Gonshor multiplication table in
this low-dimensional case ($\lambda _{111}=1$) 
\begin{eqnarray*}
\mathbf{c}_{1}^{2} &=&\lambda _{111}\mathbf{c}_{1}+\lambda _{112}\mathbf{c}%
_{2}+\lambda _{113}\mathbf{c}_{3} \\
\mathbf{c}_{1}\mathbf{c}_{2} &=&\lambda _{122}\mathbf{c}_{2}+\lambda _{123}%
\mathbf{c}_{3} \\
\mathbf{c}_{1}\mathbf{c}_{3} &=&\lambda _{133}\mathbf{c}_{3};\text{ }\mathbf{%
c}_{2}^{2}=\lambda _{223}\mathbf{c}_{3} \\
\mathbf{c}_{2}\mathbf{c}_{3} &=&\mathbf{c}_{3}^{2}=0.
\end{eqnarray*}
Here, $\lambda _{112}=-\alpha /2,$ $\lambda _{122}=\beta /2$, $\lambda
_{113}=\alpha ^{2}/16,$ $\lambda _{123}=-\beta \left( 2\alpha +\beta \right)
/16$ and $\lambda _{133}=\lambda _{223}=\beta ^{2}/8$. Considering the time
evolution $\mathbf{x}\left( t+1\right) =\mathbf{x}\left( t\right) ^{2}$ in
the Gonshor basis where $\mathbf{x}\left( t\right) =:\mathbf{c}%
_{1}+y_{2}\left( t\right) \mathbf{c}_{2}+y_{3}\left( t\right) \mathbf{c}_{3}$%
, we get 
\begin{eqnarray*}
\mathbf{x}\left( t+1\right) &=&\mathbf{c}_{1}^{2}+y_{2}^{2}\left( t\right) 
\mathbf{c}_{2}^{2}+2y_{2}\left( t\right) \mathbf{c}_{1}\mathbf{c}%
_{2}+2y_{3}\left( t\right) \mathbf{c}_{1}\mathbf{c}_{3} \\
&=&\mathbf{c}_{1}+\left( \lambda _{112}+2y_{2}\left( t\right) \lambda
_{122}\right) \mathbf{c}_{2}+\left( \lambda _{113}+2y_{2}\left( t\right)
\lambda _{123}+y_{2}^{2}\left( t\right) \lambda _{223}+2y_{3}\left( t\right)
\lambda _{133}\right) \mathbf{c}_{3} \\
&=&:y_{1}\left( t+1\right) \mathbf{c}_{1}+y_{2}\left( t+1\right) \mathbf{c}%
_{2}+y_{3}\left( t+1\right) \mathbf{c}_{3}
\end{eqnarray*}
The additional state $y_{2}^{2}\left( t\right) $ should be generated here
with $y_{2}^{2}\left( t+1\right) =\left( \lambda _{112}+2y_{2}\left(
t\right) \lambda _{122}\right) ^{2}=\lambda _{112}^{2}+4y_{2}\left( t\right)
\lambda _{112}\lambda _{122}+4y_{2}^{2}\left( t\right) \lambda _{122}^{2}$.
We obtain the closed 4-dimensional evolution 
\begin{equation*}
\left[ 
\begin{array}{c}
y_{1}\left( t+1\right) \\ 
y_{2}\left( t+1\right) \\ 
y_{2}^{2}\left( t+1\right) \\ 
y_{3}\left( t+1\right)
\end{array}
\right] =\left[ 
\begin{array}{cccc}
1 & 0 & 0 & 0 \\ 
\lambda _{112} & 2\lambda _{122} & 0 & 0 \\ 
\lambda _{112}^{2} & 4\lambda _{112}\lambda _{122} & 4\lambda _{122}^{2} & 0
\\ 
\lambda _{113} & 2\lambda _{123} & \lambda _{223} & 2\lambda _{133}
\end{array}
\right] \left[ 
\begin{array}{c}
y_{1}\left( t\right) \\ 
y_{2}\left( t\right) \\ 
y_{2}^{2}\left( t\right) \\ 
y_{3}\left( t\right)
\end{array}
\right] .
\end{equation*}
The transition matrix of the $y_{k}$s dynamics has $1$ as a dominant
eigenvalue, the corresponding eigenvector being (recalling $2\lambda
_{122}=\beta <1$ and observing $\lambda _{133}=\beta ^{2}/8<1/8$), up to a
multiplicative constant

\begin{eqnarray*}
\mathbf{y}^{\prime } &=&\left( 
\begin{array}{c}
1;\frac{\lambda _{112}}{1-2\lambda _{122}};\frac{1}{1-4\lambda _{122}^{2}}%
\left( \lambda _{112}^{2}+\frac{4\lambda _{112}^{2}\lambda _{122}}{%
1-2\lambda _{122}}\right) ; \\ 
\frac{1}{1-2\lambda _{133}}\left( \lambda _{113}+\frac{2\lambda
_{112}\lambda _{123}}{1-2\lambda _{122}}+\frac{\lambda _{223}}{1-4\lambda
_{122}^{2}}\left( \lambda _{112}^{2}+\frac{4\lambda _{112}^{2}\lambda _{122}%
}{1-2\lambda _{122}}\right) \right)
\end{array}
\right) \\
&=&:\left( 1;y_{2};y_{2}^{2};y_{3}\right) .
\end{eqnarray*}
Recalling the correspondence between the $\mathbf{x}$s and the $\mathbf{y}$%
s, namely $x_{k}=\sum_{j=1}^{3}y_{j}B_{2}\left( j,k\right) =\left( -1\right)
^{k-1}\sum_{j=k}^{3}y_{j}\binom{j-1}{k-1}$, gives the equilibrium state $%
\mathbf{x}_{eq}$ of the $\mathbf{x}$s dynamics in the simplex $\mathbf{x}%
_{eq}^{\prime }=\left( 1+y_{2}+y_{3};-y_{2}-2y_{3};y_{3}\right) ,$ with
normalizing constant $1$. For each $\alpha >0$, $0<\beta <1$, this
equilibrium point is stable because the eigenvalue $1$ is simple and
dominant. The rate at which the dynamics moves to $\mathbf{x}_{eq}$ is
geometric with parameter $2\lambda _{122}<1.$

In the boundary cases for $\left( \alpha ,\beta \right) $ for which $\alpha
=0$ and $\beta =1$, $\lambda _{112}=\lambda _{113}=0,$ $\lambda _{122}=1/2$, 
$\lambda _{123}=-1/16$ and $\lambda _{133}=\lambda _{223}=1/8$, the
transition matrix of the $y_{k}$s dynamics has $1$ as a dominant eigenvalue
with multiplicity $4$. This leads to an equilibrium quadratic skew curve of
equation 
\begin{eqnarray*}
\mathbf{x}_{eq}^{\prime } &=&\left(
y_{1}+y_{2}+y_{3};-y_{2}-2y_{3};y_{3}\right) \text{, where} \\
\text{ }y_{1} &=&1;\text{ }y_{3}=\left( y_{2}^{2}-y_{2}\right) /6.
\end{eqnarray*}
This curve is parameterized by $-2\leq y_{2}\leq 0$; it passes through the
extreme points of the simplex $\left( 0;0;1\right) $ and $\left(
1;0;0\right) $ if respectively $y_{2}=-2$ or $y_{2}=0$ and also through the
barycenter $\left( 1/3;1/3;1/3\right) $ if $y_{2}=-1$. The rate at which the
dynamics moves to the equilibrium curve $\left\{ \mathbf{x}_{eq}\right\} $
is geometric with parameter $2\lambda _{133}=1/4.$\newline

$\bullet $ \textbf{Hilbert matrices model. }With\textbf{\ }$i,j,k\geq 1$%
\textbf{, }suppose\textbf{\ } 
\begin{equation}
\gamma _{ijk}=\frac{1}{i+j-1}\text{, if }k=1,...,i+j-1\text{; }=0\text{ else.%
}  \label{B13}
\end{equation}
Note here $i,j,k$ are not bounded above by some $K$ (the model has
infinitely many species). If this is so, $\sum_{k\geq 1}\gamma _{ijk}=1$ for
all $i,j\geq 1$.

Using the change of basis $B_{2}$, with $b\left( i,j\right) =\left(
-1\right) ^{j-1}\binom{i-1}{j-1}=b^{-1}\left( i,j\right) $ and $m=i^{\prime
}+j^{\prime }-2$, we easily get that 
\begin{eqnarray*}
\lambda _{ijk} &=&b_{ii^{\prime }}b_{jj^{\prime }}\gamma _{i^{\prime
}j^{\prime }k^{\prime }}b_{k^{\prime }k}^{-1}=\sum_{i^{\prime },j^{\prime
}=1}^{i,j}\left( -1\right) ^{i^{\prime }+j^{\prime }-2}\frac{\binom{i-1}{%
i^{\prime }-1}\binom{j-1}{j^{\prime }-1}}{i^{\prime }+j^{\prime }-1}%
\sum_{k^{\prime }=k}^{i^{\prime }+j^{\prime }-1}\left( -1\right) ^{k-1}%
\binom{k^{\prime }-1}{k-1} \\
&=&\left( -1\right) ^{k-1}\sum_{m=k-1}^{i+j-2}\left( -1\right) ^{m}\frac{%
\binom{i+j-2}{m}}{m+1}\sum_{l=k-1}^{m}\binom{l}{k-1}\text{.}
\end{eqnarray*}
$\lambda _{ijk}$ depends only on $i+j$ and is $0$ if $i+j<k+1$ and also if $%
i+j>k+1$. Indeed, using the identity 
\begin{equation*}
\sum_{l=k-1}^{m}\binom{l}{k-1}=\frac{m+1-\left( k-1\right) }{k}\binom{m+1}{%
k-1},
\end{equation*}
\begin{eqnarray*}
\lambda _{ijk} &=&\frac{\left( -1\right) ^{k-1}}{k}\sum_{m=k-1}^{i+j-2}%
\left( -1\right) ^{m}\binom{i+j-2}{m}\binom{m}{k-1} \\
&=&\frac{1}{k}\binom{i+j-2}{k-1}\sum_{l=0}^{i+j-k-1}\left( -1\right) ^{l}%
\binom{i+j-k-1}{l} \\
&=&0\text{ except if }k=i+j-1.
\end{eqnarray*}

Thus $\lambda _{ijk}$ reduces to $\lambda _{ij\left( i+j-1\right) }=1/\left(
i+j-1\right) $ and $\Lambda _{k}$ is reduced to the antidiagonal $i+j=k+1.$
With $\mathbf{x}\left( t\right) =\sum_{k\geq 1}y_{k}\left( t\right) \mathbf{c%
}_{k}$, we have 
\begin{eqnarray*}
\mathbf{x}\left( t+1\right) &=&\mathbf{x}\left( t\right) ^{2}=\sum_{k\geq
1}y_{k}^{2}\left( t\right) \mathbf{c}_{k}^{2}+2\sum_{1\leq
k<l}^{{}}y_{k}^{{}}\left( t\right) y_{l}^{{}}\left( t\right) \mathbf{c}_{k}%
\mathbf{c}_{l} \\
&=&\sum_{k\geq 1}\frac{y_{k}^{2}\left( t\right) }{2k-1}\mathbf{c}%
_{2k-1}+2\sum_{1\leq k<l}^{{}}\frac{y_{k}^{{}}\left( t\right)
y_{l}^{{}}\left( t\right) }{k+l-1}\left( t\right) \mathbf{c}_{k+l-1} \\
&=&\sum_{j\geq 1}\frac{\mathbf{c}_{j}}{j}\sum_{k,l\geq
1:k+l-1=j}^{{}}y_{k}^{{}}\left( t\right) y_{l}^{{}}\left( t\right)
=:\sum_{j\geq 1}y_{j}\left( t+1\right) \mathbf{c}_{j}.
\end{eqnarray*}
so with $y_{j}\left( t+1\right) =\frac{1}{j}\sum_{k+l-1=j}^{{}}y_{k}^{{}}%
\left( t\right) y_{l}^{{}}\left( t\right) .$ To produce a triangular
infinite-dimensional linear system, we need to generate all the additional
states $y_{k}^{{}}\left( t\right) y_{l}^{{}}\left( t\right) $, $1<k<l$. For
an account on such infinite-dimensional genetic algebras, see \cite{Hol3}.%
\newline

$\bullet $ \textbf{The shift change of basis.}

We start with an example. Let $K=3$ and consider the Gonshor multiplication
table in this low-dimensional case ($\lambda _{111}=1$) 
\begin{eqnarray*}
\mathbf{c}_{1}^{2} &=&\lambda _{111}\mathbf{c}_{1}+\lambda _{112}\mathbf{c}%
_{2}+\lambda _{113}\mathbf{c}_{3} \\
\mathbf{c}_{1}\mathbf{c}_{2} &=&\lambda _{121}\mathbf{c}_{2}+\lambda _{123}%
\mathbf{c}_{3} \\
\mathbf{c}_{1}\mathbf{c}_{3} &=&\lambda _{133}\mathbf{c}_{3} \\
\mathbf{c}_{2}^{2} &=&\lambda _{223}\mathbf{c}_{3} \\
\mathbf{c}_{2}\mathbf{c}_{3} &=&\mathbf{c}_{3}^{2}=0
\end{eqnarray*}
Assume $\lambda _{ijk}>0$ and let $\mathbf{x}\left( t\right)
=\sum_{k=1}^{3}x_{k}\left( t\right) \mathbf{e}_{k}$. Then, with $\mathbf{c}%
_{1}=\mathbf{e}_{1}$, $\mathbf{c}_{2}=\mathbf{e}_{2}-\mathbf{e}_{1},$ $%
\mathbf{c}_{3}=\mathbf{e}_{3}-\mathbf{e}_{1}$, ($\mathbf{c}%
_{k}=\sum_{j}B_{1}\left( k,j\right) \mathbf{e}_{j}$), $\mathbf{x}\left(
t\right) =y_{1}\left( t\right) \mathbf{c}_{1}+y_{2}\left( t\right) \mathbf{c}%
_{2}+y_{3}\left( t\right) \mathbf{c}_{3}$ where $y_{1}\left( t\right) =1,$ $%
y_{2}\left( t\right) =x_{2}\left( t\right) $ and $y_{3}\left( t\right)
=x_{3}\left( t\right) .$ This change of basis (of type $B_{1}$) can be
inverted to give $\mathbf{e}_{1}=\mathbf{c}_{1}$, $\mathbf{e}_{2}=\mathbf{c}%
_{2}+\mathbf{c}_{1}$, $\mathbf{e}_{3}=\mathbf{c}_{3}+\mathbf{c}_{1}$. Hence, 
$x_{k}\left( t\right) =\sum_{j}y_{j}\left( t\right) B_{1}\left( j,k\right) $%
. Considering the time evolution $\mathbf{x}\left( t+1\right) =\mathbf{x}%
\left( t\right) ^{2}$ in the Gonshor canonical basis, we get 
\begin{eqnarray*}
\mathbf{x}\left( t+1\right) &=&\mathbf{c}_{1}^{2}+y_{2}^{2}\left( t\right) 
\mathbf{c}_{2}^{2}+2y_{2}\left( t\right) \mathbf{c}_{1}\mathbf{c}%
_{2}+2y_{3}\left( t\right) \mathbf{c}_{1}\mathbf{c}_{3} \\
&=&\mathbf{c}_{1}+\left( \lambda _{112}+2y_{2}\left( t\right) \lambda
_{121}\right) \mathbf{c}_{2}+\left( \lambda _{113}+\lambda
_{223}y_{2}^{2}\left( t\right) +2y_{2}\left( t\right) \lambda
_{123}+2y_{3}\left( t\right) \lambda _{133}\right) \mathbf{c}_{3} \\
&=&:y_{1}\left( t+1\right) \mathbf{c}_{1}+y_{2}\left( t+1\right) \mathbf{c}%
_{2}+y_{3}\left( t+1\right) \mathbf{c}_{3}
\end{eqnarray*}
To get a finite recursion if ever, we need to generate the evolution of the
additional state $y_{2}^{2}\left( t\right) $. We get 
\begin{equation*}
y_{2}^{2}\left( t+1\right) =\lambda _{112}^{2}+4y_{2}\left( t\right) \lambda
_{112}\lambda _{121}+4y_{2}^{2}\left( t\right) \lambda _{121}^{2}
\end{equation*}
Therefore, we obtain the closed finite-dimensional evolution 
\begin{equation*}
\left[ 
\begin{array}{l}
y_{1}\left( t+1\right) \\ 
y_{2}\left( t+1\right) \\ 
y_{2}^{2}\left( t+1\right) \\ 
y_{3}\left( t+1\right)
\end{array}
\right] =\left[ 
\begin{array}{llll}
1 & 0 & 0 & 0 \\ 
\lambda _{112} & 2\lambda _{121} & 0 & 0 \\ 
\lambda _{112}^{2} & 4\lambda _{112}\lambda _{121} & 4\lambda _{121}^{2} & 0
\\ 
\lambda _{113} & 2\lambda _{123} & \lambda _{223} & 2\lambda _{133}
\end{array}
\right] \left[ 
\begin{array}{l}
y_{1}\left( t\right) \\ 
y_{2}\left( t\right) \\ 
y_{2}^{2}\left( t\right) \\ 
y_{3}\left( t\right)
\end{array}
\right] .
\end{equation*}
The corresponding matrices $\Gamma _{k}$ given by $\gamma _{ijk}=\Gamma
_{k}\left( i,j\right) $ giving the evolution of the $x$s, are obtained while
considering the products $\mathbf{e}_{i}\mathbf{e}_{j}$ expressed in the
Gonshor basis, making use of its multiplication table and then coming back
to the natural basis. They are symmetric matrices with 
\begin{equation*}
\Gamma _{2}=\left[ 
\begin{array}{lll}
\lambda _{112} & \lambda _{112}+\lambda _{121} & \lambda _{121} \\ 
& \lambda _{112}+2\lambda _{121} & \lambda _{112}+\lambda _{121} \\ 
&  & \lambda _{112}
\end{array}
\right]
\end{equation*}
\begin{equation*}
\Gamma _{3}=\left[ 
\begin{array}{lll}
\lambda _{113} & \lambda _{113}+\lambda _{123} & \lambda _{113}+\lambda
_{133} \\ 
& \lambda _{113}+2\lambda _{123}+\lambda _{223} & \lambda _{113}+\lambda
_{123}+\lambda _{133} \\ 
&  & \lambda _{113}+2\lambda _{133}
\end{array}
\right]
\end{equation*}
\begin{equation*}
\Gamma _{1}=J-\left( \Gamma _{2}+\Gamma _{3}\right)
\end{equation*}
The entries of these matrices should be $\left[ 0,1\right] -$valued. The
compatibility conditions ensuring this (besides $\lambda _{ijk}>0$) are
found to be by inspection of the $\Gamma _{k}$s 
\begin{eqnarray*}
\max \left( 2\lambda _{121}+\lambda _{123}+\lambda _{223},\lambda
_{121}+\lambda _{123}+\lambda _{133},2\lambda _{133}\right) &\leq &1-\left(
\lambda _{112}+\lambda _{113}\right) \\
\lambda _{112}+\lambda _{113} &\leq &1.
\end{eqnarray*}
If these constraints are fulfilled (a sufficient condition being $\lambda
_{112}+\lambda _{113}+2\lambda _{121}+2\lambda _{123}+2\lambda
_{133}+\lambda _{223}\leq 1$), then the quadratic model with the above $%
\Gamma _{k}$s is Haldane linearizable along the dynamics of the $y_{k}$s.
Under the above conditions on the Gonshor structure constants, $\left( 
\mathbf{\Gamma },\mathbf{\Lambda },B_{1}\right) $ is Gonshor-compatible.

The transition matrix of the $y_{k}$s dynamics has $1$ as a dominant
eigenvalue, the corresponding eigenvector being (observing $\lambda
_{121}<1/2$ and assuming $\lambda _{133}<1/2$), up to a multiplicative
constant

\begin{eqnarray*}
\mathbf{y}^{\prime } &=&\left( 1,\frac{\lambda _{112}}{1-2\lambda _{121}}%
,\left( \frac{\lambda _{112}}{1-2\lambda _{121}}\right) ^{2},\frac{\lambda
_{113}\left( 1-2\lambda _{121}\right) ^{2}+2\lambda _{112}\lambda
_{123}\left( 1-2\lambda _{121}\right) +\lambda _{112}^{2}\lambda _{223}}{%
\left( 1-2\lambda _{121}\right) ^{2}\left( 1-2\lambda _{133}\right) }\right)
\\
&=&:\left( 1;y_{2};y_{2}^{2};y_{3}\right) .
\end{eqnarray*}
Recalling the correspondence between the $\mathbf{x}$s and the $\mathbf{y}$%
s, namely $x_{k}=\sum_{j}y_{j}B_{1}\left( j,k\right) $, we get the
equilibrium state of the $\mathbf{x}$s dynamics in the simplex $\mathbf{x}%
_{eq}^{\prime }=\left( 1-y_{2}-y_{3};y_{2};y_{3}\right) ,$ with normalizing
constant $1$. This equilibrium point is stable because the eigenvalue $1$ is
simple and dominant.

Note that in the extremal case $\lambda _{112}=\lambda _{113}=0,$ $\lambda
_{133}=1/2$, provided 
\begin{equation*}
\max \left( 2\lambda _{121}+\lambda _{123}+\lambda _{223},\lambda
_{121}+\lambda _{123}+\lambda _{133}\right) \leq 1,
\end{equation*}
$1$ is a double eigenvalue of the transition matrix for the $y_{k}$s and the
equilibrium point is $\mathbf{x}_{eq}^{\prime }=\left( 1;0;0\right) $, at
the boundary of the simplex.\newline

$\bullet $ \textbf{Gametic algebra with recombination }(\cite{Wo}, Ex.$1.3$).

Let $K=4$ and with $\theta \in \left( 0,1\right) $ and for all $i,j=1,...,4$%
, let 
\begin{equation}
\mathbf{e}_{i}\mathbf{e}_{j}=\frac{1}{2}\left( \mathbf{e}_{i}+\mathbf{e}%
_{j}\right) +\left( -1\right) ^{i\vee j-1}\frac{\theta }{2}\left( \mathbf{e}%
_{1}+\mathbf{e}_{4}-\mathbf{e}_{2}-\mathbf{e}_{3}\right) \mathbf{1}_
{\left\{ i+j=5\right\}} \label{B14}
\end{equation}
defining the $\gamma _{ijk}$s as a perturbed version of the fair Mendelian
inheritance model involving crossovers. $\theta $ is the recombination rate,
here the probability that zygote $\left( 1,4\right) $ undergoes a transition
to zygote $\left( 2,3\right) $ and conversely. In this example, with $%
\overline{\theta }:=1-\theta $%
\begin{eqnarray*}
\Gamma _{1} &=&\left[ 
\begin{array}{cccc}
1 & 1/2 & 1/2 & \overline{\theta }/2 \\ 
1/2 & 0 & \theta /2 & 0 \\ 
1/2 & \theta /2 & 0 & 0 \\ 
\overline{\theta }/2 & 0 & 0 & 0
\end{array}
\right] \text{, }\Gamma _{2}=\left[ 
\begin{array}{cccc}
0 & 1/2 & 0 & \theta /2 \\ 
1/2 & 1 & \overline{\theta }/2 & 1/2 \\ 
0 & \overline{\theta }/2 & 0 & 0 \\ 
\theta /2 & 1/2 & 0 & 0
\end{array}
\right] ,\text{ } \\
\Gamma _{3} &=&\left[ 
\begin{array}{cccc}
0 & 0 & 1/2 & \theta /2 \\ 
0 & 0 & \overline{\theta }/2 & 0 \\ 
1/2 & \overline{\theta }/2 & 1 & 1/2 \\ 
\theta /2 & 0 & 1/2 & 0
\end{array}
\right] \text{, }\Gamma _{4}=\left[ 
\begin{array}{cccc}
0 & 0 & 0 & \overline{\theta }/2 \\ 
0 & 0 & \theta /2 & 1/2 \\ 
0 & \theta /2 & 0 & 1/2 \\ 
\overline{\theta }/2 & 1/2 & 1/2 & 1
\end{array}
\right] .
\end{eqnarray*}
with $\Gamma _{1}+\Gamma _{2}+\Gamma _{3}+\Gamma _{4}=J$. And \footnote{%
Because this model is Gonshor-compatible, the Lie algebra generated by $%
\left\{ E_{1},...,E_{4}\right\} $ is solvable, see below. And all $%
E_{i}-E_{j}$ are nilpotent.} 
\begin{eqnarray*}
E_{1} &=&\left[ 
\begin{array}{cccc}
1 & 1/2 & 1/2 & \overline{\theta }/2 \\ 
0 & 1/2 & 0 & \theta /2 \\ 
0 & 0 & 1/2 & \theta /2 \\ 
0 & 0 & 0 & \overline{\theta }/2
\end{array}
\right] \text{, }E_{2}=\left[ 
\begin{array}{cccc}
1/2 & 0 & \theta /2 & 0 \\ 
1/2 & 1 & \overline{\theta }/2 & 1/2 \\ 
0 & 0 & \overline{\theta }/2 & 0 \\ 
0 & 0 & \theta /2 & 1/2
\end{array}
\right] \text{, } \\
E_{3} &=&\left[ 
\begin{array}{cccc}
1/2 & \theta /2 & 0 & 0 \\ 
0 & \overline{\theta }/2 & 0 & 0 \\ 
1/2 & \overline{\theta }/2 & 1 & 1/2 \\ 
0 & \theta /2 & 0 & 1/2
\end{array}
\right] \text{, }E_{4}=\left[ 
\begin{array}{cccc}
\overline{\theta }/2 & 0 & 0 & 0 \\ 
\theta /2 & 1/2 & 0 & 0 \\ 
\theta /2 & 0 & 1/2 & 0 \\ 
\overline{\theta }/2 & 1/2 & 1/2 & 1
\end{array}
\right] .
\end{eqnarray*}
Using, 
\begin{equation*}
B_{4}:=\left[ 
\begin{array}{llll}
1 &  &  &  \\ 
1 & -1 &  &  \\ 
1 & 0 & -1 &  \\ 
1 & -1 & -1 & 1
\end{array}
\right] ,\text{ with }B_{4}^{-1}=B_{4},\text{ we get}
\end{equation*}
\begin{eqnarray*}
\mathbf{c}_{1}^{2} &=&\lambda _{111}\mathbf{c}_{1}=\mathbf{c}_{1} \\
\mathbf{c}_{1}\mathbf{c}_{2} &=&\mathbf{c}_{2}/2\text{, }\mathbf{c}_{1}%
\mathbf{c}_{3}=\mathbf{c}_{3}/2 \\
\mathbf{c}_{1}\mathbf{c}_{4} &=&\left( 1-\theta \right) \mathbf{c}_{4}/2%
\text{, }\mathbf{c}_{2}^{2}=0\text{, }\mathbf{c}_{2}\mathbf{c}_{3}=\theta 
\mathbf{c}_{4}/2 \\
\mathbf{c}_{2}\mathbf{c}_{4} &=&\mathbf{c}_{3}^{2}=\mathbf{c}_{3}\mathbf{c}%
_{4}=\mathbf{c}_{4}^{2}=0
\end{eqnarray*}
which is Gonshor-like with right train roots $\lambda _{111}=1,$ $\lambda
_{122}=\lambda _{133}=1/2$, $\lambda _{144}=\left( 1-\theta \right) /2$.
Because $\lambda _{122}=1/2$ is a train root with multiplicity $2$, we
expect an equilibrium surface for this model, (\cite{Gon2}, \cite{Lyu}\emph{%
, theorem 7.2.6)}. Considering indeed the time evolution $\mathbf{x}\left(
t+1\right) =\mathbf{x}\left( t\right) ^{2}$ in the Gonshor basis where $%
\mathbf{x}\left( t\right) =:\mathbf{c}_{1}+y_{2}\left( t\right) \mathbf{c}%
_{2}+y_{3}\left( t\right) \mathbf{c}_{3}+y_{4}\left( t\right) \mathbf{c}_{4}$%
, we get 
\begin{eqnarray*}
\mathbf{x}\left( t+1\right) &=&\mathbf{c}_{1}+y_{2}\left( t\right) \mathbf{c}%
_{2}+y_{3}\left( t\right) \mathbf{c}_{3}+\left( \left( 1-\theta \right)
y_{4}\left( t\right) +\theta y_{2}\left( t\right) y_{3}\left( t\right)
\right) \mathbf{c}_{4} \\
&=&:y_{1}\left( t+1\right) \mathbf{c}_{1}+y_{2}\left( t+1\right) \mathbf{c}%
_{2}+y_{3}\left( t+1\right) \mathbf{c}_{3}+y_{4}\left( t+1\right) \mathbf{c}%
_{4}.
\end{eqnarray*}
To get a finite recursion, we need to generate the evolution of one
additional state, namely $y_{2}\left( t\right) y_{3}\left( t\right) $. We
simply get 
\begin{equation*}
y_{2}\left( t+1\right) y_{3}\left( t+1\right) =y_{2}\left( t\right)
y_{3}\left( t\right) .
\end{equation*}
We obtain the closed finite-dimensional evolution 
\begin{equation*}
\left[ 
\begin{array}{c}
y_{1}\left( t+1\right) \\ 
y_{2}\left( t+1\right) \\ 
y_{3}\left( t+1\right) \\ 
y_{2}y_{3}\left( t+1\right) \\ 
y_{4}\left( t+1\right)
\end{array}
\right] =\left[ 
\begin{array}{ccccc}
1 &  &  &  &  \\ 
0 & 1 &  &  &  \\ 
0 & 0 & 1 &  &  \\ 
0 & 0 & 0 & 1 &  \\ 
0 & 0 & 0 & \theta & 1-\theta
\end{array}
\right] \left[ 
\begin{array}{c}
y_{1}\left( t\right) \\ 
y_{2}\left( t\right) \\ 
y_{3}\left( t\right) \\ 
y_{2}y_{3}\left( t\right) \\ 
y_{4}\left( t\right)
\end{array}
\right] .
\end{equation*}
The transition matrix of this dynamics has $1$ as a dominant eigenvalue with
multiplicity $4$, the corresponding eigenvector being, up to two
indeterminate constants $y_{2},y_{3}$: $\left(
1,y_{2},y_{3},y_{2}y_{3},y_{2}y_{3}\right) .$

Recalling the correspondence between the $\mathbf{x}$s and the $\mathbf{y}$%
s, namely $x_{k}=\sum_{j=1}^{4}y_{j}B_{4}\left( j,k\right) $, in view of $%
\mathbf{y}_{*}^{\prime }=\left( 1,y_{2},y_{3},y_{2}y_{3}\right) $, leads to
equilibrium states $\mathbf{x}_{eq}$ of the $\mathbf{x}$s dynamics in the
simplex given by $\mathbf{x}_{eq}^{\prime }=\left(
1+y_{2}+y_{3}+y_{2}y_{3};-y_{2}-y_{2}y_{3};-y_{3}-y_{2}y_{3};y_{2}y_{3}%
\right) ,$ with normalizing constant $1$ and for those values of $-1\leq
y_{2},y_{3}\leq 0$ for which $\mathbf{x}_{eq}$ belongs to the simplex. This
equilibrium hypervolume, parameterized by $y_{2},y_{3}$, is skew; the
equilibrium surface is defined as the intersection of the simplex $S_{4}$
with the latter hypervolume which is seen to be of equation $%
x_{2}x_{3}=x_{1}x_{4}$. It is stable and the rate at which the dynamics
moves to the equilibrium surface $\left\{ \mathbf{x}_{eq}\right\} $ is
geometric with parameter $1-\theta <1$. Note that $\left\{ \mathbf{x}%
_{eq}\right\} $\textbf{\ }contains the faces of the simplex: $\left(
0;1+y_{3};0;-y_{3}\right) $ and $\left( 0;0;1+y_{2};-y_{2}\right) $ obtained
respectively when $y_{2}=-1$ and $y_{3}=-1$, together with the barycenter of
the simplex obtained when $y_{2}=y_{3}=-1/2$. Coming back to the natural
basis, it can be checked in addition that in this example 
\begin{equation*}
\mathbf{x}\left( t+1\right) -\mathbf{x}\left( t\right) =\theta \left(
x_{2}\left( t\right) x_{3}\left( t\right) -x_{1}\left( t\right) x_{4}\left(
t\right) \right) \mathbf{u,}
\end{equation*}
where $\mathbf{u}^{\prime }=\left( 1,-1,-1,1\right) $. So $\mathbf{x}\left(
t\right) $ moves in the direction of $\mathbf{u}$, starting from $\mathbf{x}%
\left( 0\right) $, before hitting the set $\left\{ \mathbf{x}_{eq}\right\} $%
: the domain of attraction of a point in $\left\{ \mathbf{x}_{eq}\right\} $
is included in a line pointing to $\left\{ \mathbf{x}_{eq}\right\} $ from $%
\mathbf{x}\left( 0\right) $ in the direction of $\mathbf{u}$.

\subsection{Models not in the class of genetic algebras}

So far, we gave some examples of symmetric matrices $\Gamma _{k}$ (obeying $%
\forall i,j,$ $\sum_{k}\Gamma _{k}\left( i,j\right) =1$) leading to genetic
algebras which are linearizable in higher dimension. We now give some
examples which are not. From the previous arguments, the necessary and
sufficient conditions under which a choice of $\Gamma _{i}$ leads to genetic
algebras is that:

1/ The matrices $E_{i}$ with $E_{i}\left( k,j\right) =\Gamma _{k}\left(
i,j\right) =\gamma _{ijk}$ should be simultaneously triangularizable (ST) and

2/ $\forall i<j,$ $E_{i}-E_{j}$ should be nilpotent matrices.\newline

A particular stochastic model $\left\{ \mathbf{\Gamma }\right\} $ may fail
to be Gonshor-compatible if condition 1/ or 2/ or both fail.\newline

Concerning condition 1/: Quasi-commutative matrices are matrices commuting
with their commutators (with commuting matrices being quasi-commutative). If 
$\forall i<j,\forall k,$ $\left[ E_{k},\left[ E_{i},E_{j}\right] \right] =0$%
, then the set of matrices $E_{i}$ are said to be quasi-commutative and in
this case the $E_{i}$ are simultaneously triangularizable (ST) in the
extension $\Bbb{C}$ of $\Bbb{R}$, \cite{MC1}. Commuting matrices are even
simultaneously diagonalizable.

If quasi-commutativity is a ST sufficient condition, it is not necessary. In 
\cite{MC2}, the necessary and sufficient condition for ST was shown to be: $%
\forall i<j$, $P\left( E_{1},...,E_{K}\right) \left[ E_{i},E_{j}\right] $
are nilpotent matrices for any polynomial $P$ in the possibly
non-commutative variables $\left\{ E_{1},...,E_{K}\right\} $. By Theorem $3$
in \cite{MC2}, this condition is equivalent to the solvability of the Lie
algebra $L:=\left\langle E_{1},...,E_{K}\right\rangle _{LA}$ spanned by $%
\left\{ E_{1},...,E_{K}\right\} $, closing the linear space generated by the 
$E_{i}$ with respect to the commutator operation (the solvability of $L$
means that its derived series terminates in the zero subalgebra \footnote{%
This means that for all examples designed in Section $3.1$, the Lie algebras
generated by the $\left\{ E_{i}\right\} $ which can be built from the $%
\left\{ \Gamma _{i}\right\} $ we started from, were solvable and that all $%
E_{i}-E_{j}$ were nilpotent.}). $L$ has $d$ ($K\leq $dim$\left( L\right)
=d\leq K^{2}$) linearly independent basis matrices $\left\{
E_{1},...,E_{d}\right\} $, with $\left\{ E_{1},...,E_{K}\right\} \subseteq
\left\{ E_{1},...,E_{d}\right\} $, and $\left[ E_{i},E_{j}\right]
=\sum_{k=1}^{d}e_{ijk}E_{k}$ where $e_{ijk}$ are the structure constants of $%
L$ obeying $e_{ijk}=-e_{jik}$ and the Jacobi identity. The matrix $K$
associated to the Killing form of $L$ is $K:=\left[ k_{i,j}\right] $, where $%
k_{i,j}=\sum_{k,l}e_{ilk}e_{jkl}$. Its non-degeneracy is a signature of the
semi-simplicity of $L$, with semi-simplicity$\Rightarrow $non-solvability
(the reciprocal being false in general).

A constructive (although prohibitive even for small $K$) test for pair-wise
ST of $\left\{ E_{1},...,E_{K}\right\} $ is that of Theorem $6$ of \cite{Alp}%
: for every $k$ $\in \left\{ 1,K^{2}-1\right\} $, $\forall i<j$, with $%
U_{l}\in \left\{ E_{i},E_{j}\right\} $, $l=1,...,k$, each matrix of the form 
$U_{1}\cdot \cdot \cdot U_{k}\left[ E_{i},E_{j}\right] $ has zero trace. ST
of $\left\{ E_{1},...,E_{K}\right\} $ condition is: for every $k$ $\in
\left\{ 1,K^{K}-1\right\} $, $\forall i<j$, with $U_{l}\in \left\{
E_{1},...,E_{K}\right\} $, $l=1,...,k$, each matrix of the form $U_{1}\cdot
\cdot \cdot U_{k}\left[ E_{i},E_{j}\right] $ has zero trace.\newline

The conditions 1/ and 2/ can be used to show that special important families
of $\Gamma _{k}$ do not lead to genetic algebras.\newline

$\bullet $\textbf{\ Permutations.}

$\left( i\right) $ Suppose 
\begin{equation*}
\Gamma _{1}=\left[ 
\begin{array}{cc}
0 & 1 \\ 
1 & 0
\end{array}
\right] \text{, }\Gamma _{2}=\left[ 
\begin{array}{cc}
1 & 0 \\ 
0 & 1
\end{array}
\right] .
\end{equation*}
Then $E_{1}=\Gamma _{1}$ and $E_{2}=\Gamma _{2}$ are commuting matrices so
simultaneously triangularizable (in fact diagonalizable). However $%
E_{1}-E_{2}=\left[ 
\begin{array}{cc}
-1 & 1 \\ 
1 & -1
\end{array}
\right] $ with trace $-2$ is not nilpotent.\newline

$\left( ii\right) $ Suppose 
\begin{equation*}
\Gamma _{1}=\left[ 
\begin{array}{ccc}
0 & 1 & 0 \\ 
1 & 0 & 0 \\ 
0 & 0 & 1
\end{array}
\right] \text{, }\Gamma _{2}=\left[ 
\begin{array}{ccc}
0 & 0 & 1 \\ 
0 & 1 & 0 \\ 
1 & 0 & 0
\end{array}
\right] \text{, }\Gamma _{3}=\left[ 
\begin{array}{ccc}
1 & 0 & 0 \\ 
0 & 0 & 1 \\ 
0 & 1 & 0
\end{array}
\right] .
\end{equation*}
Then 
\begin{equation*}
E_{1}=\left[ 
\begin{array}{ccc}
0 & 1 & 0 \\ 
0 & 0 & 1 \\ 
1 & 0 & 0
\end{array}
\right] \text{, }E_{2}=\left[ 
\begin{array}{ccc}
1 & 0 & 0 \\ 
0 & 1 & 0 \\ 
0 & 0 & 1
\end{array}
\right] \text{, }E_{3}=\left[ 
\begin{array}{ccc}
0 & 0 & 1 \\ 
1 & 0 & 0 \\ 
0 & 1 & 0
\end{array}
\right]
\end{equation*}
which are commuting permutation matrices so simultaneously triangularizable
(in fact diagonalizable with a unitary matrix). However $E_{1}-E_{2}=\left[ 
\begin{array}{ccc}
-1 & 1 & 0 \\ 
0 & -1 & 1 \\ 
1 & 0 & -1
\end{array}
\right] $ with trace $-3$ is not nilpotent.\newline

$\left( iii\right) $ Suppose 
\begin{eqnarray*}
\Gamma _{1} &=&\left[ 
\begin{array}{cccc}
0 & 1 & 0 & 0 \\ 
1 & 0 & 0 & 0 \\ 
0 & 0 & 0 & 1 \\ 
0 & 0 & 1 & 0
\end{array}
\right] \text{, }\Gamma _{2}=\left[ 
\begin{array}{cccc}
0 & 0 & 1 & 0 \\ 
0 & 1 & 0 & 0 \\ 
1 & 0 & 0 & 0 \\ 
0 & 0 & 0 & 1
\end{array}
\right] \text{, } \\
\Gamma _{3} &=&\left[ 
\begin{array}{cccc}
1 & 0 & 0 & 0 \\ 
0 & 0 & 0 & 1 \\ 
0 & 0 & 1 & 0 \\ 
0 & 1 & 0 & 0
\end{array}
\right] \text{, }\Gamma _{4}=\left[ 
\begin{array}{cccc}
0 & 0 & 0 & 1 \\ 
0 & 0 & 1 & 0 \\ 
0 & 1 & 0 & 0 \\ 
1 & 0 & 0 & 0
\end{array}
\right] .
\end{eqnarray*}
with $\Gamma _{1}+\Gamma _{2}+\Gamma _{3}+\Gamma _{4}=J$. Then 
\begin{eqnarray*}
E_{1} &=&\left[ 
\begin{array}{cccc}
0 & 1 & 0 & 0 \\ 
0 & 0 & 1 & 0 \\ 
1 & 0 & 0 & 0 \\ 
0 & 0 & 0 & 1
\end{array}
\right] \text{, }E_{2}=\left[ 
\begin{array}{cccc}
1 & 0 & 0 & 0 \\ 
0 & 1 & 0 & 0 \\ 
0 & 0 & 0 & 1 \\ 
0 & 0 & 1 & 0
\end{array}
\right] \text{, } \\
E_{3} &=&\left[ 
\begin{array}{cccc}
0 & 0 & 0 & 1 \\ 
1 & 0 & 0 & 0 \\ 
0 & 0 & 1 & 0 \\ 
0 & 1 & 0 & 0
\end{array}
\right] \text{, }E_{4}=\left[ 
\begin{array}{cccc}
0 & 0 & 1 & 0 \\ 
0 & 0 & 0 & 1 \\ 
0 & 1 & 0 & 0 \\ 
1 & 0 & 0 & 0
\end{array}
\right]
\end{eqnarray*}
are also permutation (non-symmetric) matrices which are not even
quasi-commuting. Note $E_{1}+E_{2}+E_{3}+E_{4}=J$. We have for instance 
\begin{equation*}
E_{2}E_{3}\left[ E_{1},E_{2}\right] =\left[ 
\begin{array}{cccc}
-1 & 0 & 1 & 0 \\ 
0 & 0 & 0 & 0 \\ 
0 & 0 & -1 & 1 \\ 
1 & 0 & 0 & -1
\end{array}
\right]
\end{equation*}
with trace $-3$, so not nilpotent. Moreover, 
\begin{equation*}
E_{1}-E_{2}=\left[ 
\begin{array}{cccc}
-1 & 1 & 0 & 0 \\ 
0 & -1 & 1 & 0 \\ 
1 & 0 & 0 & -1 \\ 
0 & 0 & -1 & 1
\end{array}
\right] ,
\end{equation*}
with trace $-1$, is not nilpotent. The (non-solvable) Lie algebra $L$
generated by the $E_{i}$, $i=1,...,4$, has dimension $d=10$, with basis 
\begin{equation*}
\begin{array}{c}
\{E_{1};E_{2};E_{3};E_{4};E_{5}=\left[ E_{1},E_{2}\right] ;E_{6}=\left[
E_{1},E_{3}\right] ; \\ 
E_{7}=\left[ E_{1},E_{5}\right] ;E_{8}=\left[ E_{1},E_{6}\right]
;E_{9}=\left[ E_{1},E_{8}\right] ;E_{10}=\left[ E_{2},E_{5}\right] \}.
\end{array}
\end{equation*}
The associated structure constants can be computed, together with the
associated Killing matrix $K$ which is found to be of rank $8$, so
degenerate. The Lie algebra $L$ is neither solvable nor semisimple.\newline

$\left( iii^{\prime }\right) $ Suppose 
\begin{equation*}
\Gamma _{1}=I\text{, }\Gamma _{2}=\left[ 
\begin{array}{cccc}
0 & 1 & 0 & 0 \\ 
1 & 0 & 0 & 0 \\ 
0 & 0 & 0 & 1 \\ 
0 & 0 & 1 & 0
\end{array}
\right] ,\text{ }\Gamma _{3}=\left[ 
\begin{array}{cccc}
0 & 0 & 1 & 0 \\ 
0 & 0 & 0 & 1 \\ 
1 & 0 & 0 & 0 \\ 
0 & 1 & 0 & 0
\end{array}
\right] \text{, }\Gamma _{4}=\left[ 
\begin{array}{cccc}
0 & 0 & 0 & 1 \\ 
0 & 0 & 1 & 0 \\ 
0 & 1 & 0 & 0 \\ 
1 & 0 & 0 & 0
\end{array}
\right] .
\end{equation*}
with $\Gamma _{1}+\Gamma _{2}+\Gamma _{3}+\Gamma _{4}=J$. Then $E_{i}=\Gamma
_{i}$, $i=1,...,4$ are also permutation (symmetric) matrices which are
commuting (the Lie algebra $L$ generated by the $E_{i}$ is solvable of order 
$1$). However, 
\begin{equation*}
E_{1}-E_{2}=\left[ 
\begin{array}{cccc}
1 & -1 & 0 & 0 \\ 
-1 & 1 & 1 & 0 \\ 
1 & 0 & 1 & -1 \\ 
0 & 0 & -1 & 1
\end{array}
\right] ,
\end{equation*}
with trace $4$, is not nilpotent.\newline

These examples suggest that, would $\Gamma _{k}$ be symmetric (involutive)
permutation matrices, such models should not lead to genetic algebras in
general (Recall though that the fixed equilibrium point of such dynamics is
always the barycenter $\mathbf{x}_{B}$ of the simplex $S_{K}$). This
suggestion is not reduced to symmetric permutation matrices. Suppose 
\begin{equation*}
\Gamma _{1}=\left[ 
\begin{array}{ccc}
0 & 1/2 & 1/2 \\ 
1/2 & 0 & 1/2 \\ 
1/2 & 1/2 & 0
\end{array}
\right] =\Gamma _{2},\text{ }\Gamma _{3}=I,
\end{equation*}
the symmetrized version of the non-symmetric permutation matrices 
\begin{equation*}
P_{1}=\left[ 
\begin{array}{ccc}
0 & 1 & 0 \\ 
0 & 0 & 1 \\ 
1 & 0 & 0
\end{array}
\right] \text{, }P_{2}=\left[ 
\begin{array}{ccc}
0 & 0 & 1 \\ 
1 & 0 & 0 \\ 
0 & 1 & 0
\end{array}
\right] ,\text{ }P_{3}=I.
\end{equation*}
Then 
\begin{equation*}
E_{1}=\left[ 
\begin{array}{ccc}
0 & 1/2 & 1/2 \\ 
0 & 1/2 & 1/2 \\ 
1 & 0 & 0
\end{array}
\right] \text{, }E_{2}=\left[ 
\begin{array}{ccc}
1/2 & 0 & 1/2 \\ 
1/2 & 0 & 1/2 \\ 
0 & 1 & 0
\end{array}
\right] \text{, }E_{3}=\left[ 
\begin{array}{ccc}
1/2 & 1/2 & 0 \\ 
1/2 & 1/2 & 0 \\ 
0 & 0 & 1
\end{array}
\right]
\end{equation*}
which are non-quasi-commuting bistochastic matrices, however with $%
E_{1}\left[ E_{1},E_{2}\right] $, $E_{3}\left[ E_{1},E_{2}\right] $
nilpotent for instance. But $E_{1}-E_{3}=\left[ 
\begin{array}{ccc}
-1/2 & 0 & 1/2 \\ 
-1/2 & 0 & 1/2 \\ 
1 & 0 & -1
\end{array}
\right] $ with trace $-3/2$ is not nilpotent. The Lie algebra $L$ generated
by the $E_{i}$, $i=1,...,3$, has dimension $d=3$, with basis $%
\{E_{1};E_{2};E_{3}\}$. The associated structure constants can be computed,
together with the associated Killing matrix $K$ which is found to be of rank 
$1$, so degenerate. The Lie algebra $L$ is solvable of order $2$ (the
brackets $\left[ E_{i},E_{j}\right] $, $i<j$ being proportional to the same
matrix $E_{1}-E_{2}$) and not semisimple.\newline

$\bullet $ \textbf{The general }$2-$\textbf{dimensional stochastic case,
including the} \textbf{bistochastic matrices case.}

$\left( iv\right) $ With $\alpha ,\beta ,\gamma \in \left( 0,1\right) $ and $%
\overline{\alpha }=1-\alpha ,$ $\overline{\beta }=1-\beta ,$ $\overline{%
\gamma }=1-\gamma $, suppose 
\begin{equation*}
\Gamma _{1}=\left[ 
\begin{array}{cc}
\alpha & \beta \\ 
\beta & \gamma
\end{array}
\right] \text{, }\Gamma _{2}=\left[ 
\begin{array}{cc}
\overline{\alpha } & \overline{\beta } \\ 
\overline{\beta } & \overline{\gamma }
\end{array}
\right] ,
\end{equation*}
the general $2-$dimensional stochastic problem. Then 
\begin{equation*}
E_{1}=\left[ 
\begin{array}{cc}
\alpha & \beta \\ 
\overline{\alpha } & \overline{\beta }
\end{array}
\right] \text{, }E_{2}=\left[ 
\begin{array}{cc}
\beta & \gamma \\ 
\overline{\beta } & \overline{\gamma }
\end{array}
\right] ,
\end{equation*}
which do not commute in general (unless $\beta \overline{\beta }=\gamma 
\overline{\alpha }$). The Lie algebra $L$ generated by $\left\{
E_{1},E_{2}\right\} $ has dimension $d=3$ with basis $\left\{
E_{1},E_{2},E_{3}=\left[ E_{1},E_{2}\right] \right\} $ if $\alpha +\gamma
\neq 2\beta $ and dimension $d=2$ if $\alpha +\gamma =2\beta $. It is
solvable in both cases because, by Cartan solvability criterion, the Killing
form $K$ satisfies $K(E,E^{\prime })=$Trace$\left( E,E^{\prime }\right) =0$
for all $E$ in $L$ and $E^{\prime }$ in $[L,L]$. However here, $%
E_{1}-E_{2}=\left[ 
\begin{array}{cc}
\alpha -\beta & \beta -\gamma \\ 
\overline{\alpha }-\overline{\beta } & \overline{\beta }-\overline{\gamma }
\end{array}
\right] ,$ with trace $\alpha -\beta +\overline{\beta }-\overline{\gamma }.$
It is not nilpotent unless $\alpha +\gamma =2\beta $. Although $L$ is
solvable, the general $2-$dimensional stochastic problem is not
Gonshor-linearizable unless $\alpha +\gamma =2\beta $.

If $\alpha +\gamma =2\beta $, the evolutionary dynamics $\mathbf{x}\left(
t+1\right) =\mathbf{x}\left( t\right) ^{2}$ reads 
\begin{eqnarray*}
x_{1}\left( t+1\right) &=&\left( x_{1}\left( t\right) ,1-x_{1}\left(
t\right) \right) \Gamma _{1}\left( x_{1}\left( t\right) ,1-x_{1}\left(
t\right) \right) ^{\prime }=2\left( \beta -\gamma \right) x_{1}\left(
t\right) +\gamma \\
x_{2}\left( t+1\right) &=&2\left( \beta -\gamma \right) x_{2}\left( t\right)
+\overline{\alpha }.
\end{eqnarray*}
It is indeed linear with fixed point $\mathbf{x}_{eq}=\left( \gamma /\left(
\gamma +\overline{\alpha }\right) ;\overline{\alpha }/\left( \gamma +%
\overline{\alpha }\right) \right) ^{\prime }$, in the simplex. So, except in
this particular case, the general $2-$dimensional problem is not amenable to
a linear problem and when it is, there is no additional state to generate.

However, because of the very low dimension ($K=2$) of the problem, the
analysis of the model with $\left\{ \Gamma _{1},\Gamma _{2}\right\} $
defined above is possible. We find that for any $\alpha ,\beta ,\gamma \in
\left( 0,1\right) $, the $2-$dimensional dynamics $x_{k}\left( t+1\right) =%
\mathbf{x}^{\prime }\Gamma _{k}\mathbf{x}$, $k=1,..,2$ always has a fixed
point in the simplex. Defining $\varepsilon =\left( \alpha +\gamma \right)
/2-\beta $, the dynamics is 
\begin{equation}
x_{1}\left( t+1\right) =2\varepsilon x_{1}\left( t\right) ^{2}+\left( \alpha
-\gamma -2\varepsilon \right) x_{1}\left( t\right) +\gamma =:f\left(
x_{1}\left( t\right) \right) \text{,}  \label{B15}
\end{equation}
with a quadratic $f$. With $\Delta =\left( 2\beta -1\right) ^{2}+4\gamma 
\overline{\alpha }>0$, the fixed point in the simplex therefore is 
\begin{equation*}
x_{1,eq}=\frac{\gamma -\alpha +2\varepsilon +1-\sqrt{\Delta }}{4\varepsilon }%
\text{, }x_{2,eq}=1-x_{1,eq}\text{.}
\end{equation*}
With $\Delta >1\Leftrightarrow \beta \overline{\beta }<\gamma \overline{%
\alpha }$, we have $f^{\prime }\left( x_{1,eq}\right) =1-\sqrt{\Delta }$
with $\left| f^{\prime }\left( x_{1,eq}\right) \right| <1$ if $\Delta \leq 1$
or $1<\Delta \leq 4$. So $x_{1,eq}$ is asymptotically stable if and only if $%
\Delta \leq 4$. If $\Delta >4$, $x_{1}\left( t\right) $ oscillates between
two limiting values in the simplex around $x_{1,eq}$, as a center fixed and
unstable point: we have two period-two equilibrium points (obeying $f\left(
f\left( x\right) \right) =x$). If $4>\Delta >1$, $x_{1}\left( t\right) $
tends to $x_{1,eq}$ while oscillating around $x_{1,eq}$, as a fixed stable
equilibrium point. Else, if $\Delta \leq 1$, $x_{1}\left( t\right) $ tends
to $x_{1,eq}$ from below or from above (depending on the initial condition)
without over-crossing its limiting value more than once.\newline

If $\beta =\overline{\alpha }=\overline{\gamma }\Rightarrow \beta \overline{%
\beta }=\gamma \overline{\alpha }$, $\Gamma _{1}$ and $\Gamma _{2}$ are
bistochastic and $\left[ E_{1},E_{2}\right] =0$. In this case, $%
E_{1}-E_{2}=\left[ 
\begin{array}{cc}
2\alpha -1 & 1-2\alpha \\ 
1-2\alpha & 2\alpha -1
\end{array}
\right] $ with zero trace. This matrix is not nilpotent unless $\alpha
=\beta =\gamma =1/2$, a trivial case. In dimension $K=2$, bistochastic
models are not Gonshor-linearizable in general either.\newline

$\bullet $ \textbf{Unbalanced Mendelian inheritance model.}

$\left( v\right) $ With $a_{1}+a_{2}=1,$ $b_{1}+b_{3}=1,$ $b_{2}+a_{3}=1,$
suppose ($\Gamma _{1}+\Gamma _{2}+\Gamma _{3}=J$) 
\begin{equation*}
\Gamma _{1}=\left[ 
\begin{array}{ccc}
1 & a_{1} & b_{1} \\ 
a_{1} & 0 & 0 \\ 
b_{1} & 0 & 0
\end{array}
\right] \text{, }\Gamma _{2}=\left[ 
\begin{array}{ccc}
0 & a_{2} & 0 \\ 
a_{2} & 1 & b_{2} \\ 
0 & b_{2} & 0
\end{array}
\right] ,\text{ }\Gamma _{3}=\left[ 
\begin{array}{ccc}
0 & 0 & b_{3} \\ 
0 & 0 & a_{3} \\ 
b_{3} & a_{3} & 1
\end{array}
\right] .
\end{equation*}
This is a model with Mendelian segregation for which only interactions $%
\left( k,j\right) $ or $\left( i,k\right) $ can produce type-$k$ offspring.
Here $\mathbf{x}\left( t+1\right) =\mathbf{x}\left( t\right) ^{2}$ where $%
\mathbf{x}\left( t\right) =\sum_{i}x_{i}\left( t\right) \mathbf{e}_{i}$ and
multiplication table given by $\mathbf{e}_{i}\mathbf{e}_{j}=\Gamma
_{i}\left( i,j\right) \mathbf{e}_{i}+\Gamma _{j}\left( i,j\right) \mathbf{e}%
_{j}$, where $\Gamma _{i}\left( i,j\right) +\Gamma _{j}\left( i,j\right) =1$%
. As observed previously in Section $2$, the dynamics of species frequencies
is also $x_{k}\left( t+1\right) =\mathbf{x}^{\prime }\Gamma _{k}\mathbf{x}$, 
$k=1,..,3$. It can alternatively be written in vector form as 
\begin{equation}
\mathbf{x}\left( t+1\right) =\mathbf{x}\left( t\right) +D_{\mathbf{x}\left(
t\right) }A\mathbf{x}\left( t\right) ,  \label{B16}
\end{equation}
where $A:=\left[ 
\begin{array}{ccc}
0 & 2a_{1}-1 & 2b_{1}-1 \\ 
2a_{2}-1 & 0 & 2b_{2}-1 \\ 
2b_{3}-1 & 2a_{3}-1 & 0
\end{array}
\right] $ is a skew-symmetric matrix. In such a case, 
\begin{equation*}
E_{1}=\left[ 
\begin{array}{ccc}
1 & a_{1} & b_{1} \\ 
0 & a_{2} & 0 \\ 
0 & 0 & b_{3}
\end{array}
\right] \text{, }E_{2}=\left[ 
\begin{array}{ccc}
a_{1} & 0 & 0 \\ 
a_{2} & 1 & b_{2} \\ 
0 & 0 & a_{3}
\end{array}
\right] \text{, }E_{3}=\left[ 
\begin{array}{ccc}
b_{1} & 0 & 0 \\ 
0 & b_{2} & 0 \\ 
b_{3} & a_{3} & 1
\end{array}
\right]
\end{equation*}
which are non-commuting column stochastic matrices with $E_{1}\left[
E_{1},E_{2}\right] $, $E_{2}\left[ E_{1},E_{2}\right] $, $E_{1}E_{2}\left[
E_{1},E_{2}\right] $,..., nilpotent matrices.

However $E_{1}-E_{2}=\left[ 
\begin{array}{ccc}
1-a_{1} & a_{1} & b_{1} \\ 
-a_{2} & a_{2}-1 & -b_{2} \\ 
0 & 0 & b_{3}-a_{3}
\end{array}
\right] $ with trace $a_{2}-a_{1}+b_{3}-a_{3}$ is not nilpotent unless $%
a_{1}=a_{2}=1/2$, $b_{3}=a_{3}$ and $b_{1}=b_{2}.$ Similarly, $%
E_{2}-E_{3}=\left[ 
\begin{array}{ccc}
a_{1}-b_{1} & 0 & 0 \\ 
a_{2} & 1-b_{2} & b_{2} \\ 
-b_{3} & -a_{3} & a_{3}-1
\end{array}
\right] $ with trace $a_{1}-b_{1}+a_{3}-b_{2}$ is not nilpotent unless $%
a_{3}=b_{2}=1/2$, $b_{1}=a_{1}$, and $a_{2}=b_{3}$ and $E_{1}-E_{3}=\left[ 
\begin{array}{ccc}
1-b_{1} & a_{1} & b_{1} \\ 
0 & a_{2}-b_{2} & 0 \\ 
-b_{3} & -a_{3} & b_{3}-1
\end{array}
\right] $ is not nilpotent unless $a_{2}=b_{2}$, $b_{1}=b_{3}=1/2$, and $%
a_{1}=a_{3}.$ This shows that the only case when $E_{i}-E_{j}$ are all
nilpotent is the trivial balanced (fair) Mendelian case when $%
a_{1}=a_{2}=a_{3}=b_{1}=b_{2}=b_{3}=1/2$, corresponding to $A=0$ with $%
\mathbf{x}\left( t+1\right) =\mathbf{x}\left( t\right) $, its linear but
uninteresting corresponding dynamics. This suggests that unbalanced
Mendelian segregation dynamics should not be Gonshor-linearizable in general.%
\newline

\textbf{Acknowledgments:}

T. Huillet acknowledges support from the \textit{Project Basal PFB 03} of
the CONICYT of Chile, from the ``Chaire \textit{Mod\'{e}lisation
math\'{e}matique et biodiversit\'{e}''} and, together with N. Grosjean, from
the labex MME-DII Center of Excellence (\textit{Mod\`{e}les
math\'{e}matiques et \'{e}conomiques de la dynamique, de l'incertitude et
des interactions}, ANR-11-LABX-0023-01 project).


\begin{thebibliography}{99}
\bibitem{Ab1}  Abraham, V. M. Linearizing quadratic transformations in
genetic algebras. Proc London Math Soc 40: 346-363, 1980.

\bibitem{Ab2}  Abraham, V. M. The genetic algebra of polyploids. Proc.
London Math. Soc. (3) 40, 385-429, 1980.

\bibitem{ACL}  Andrade R.; Catalan A.; Labra A. The identity $\left(
x^{2}\right) ^{2}=\varpi \left( x\right) x^{3}$ in baric algebras. In:
Non-Associative Algebra and Its Applications. Math. and its Applic., S.
Gonzalez, Ed., Springer Science+Business Media, B.V., 1994.

\bibitem{Alp}  Al'pin Yu. A.; Koreshkov, N. A. On the Simultaneous
Triangulability of Matrices. Mathematical Notes, Vol. 68, No. 5, 2000.

\bibitem{Burg}  B\"{u}rger, R. \emph{The mathematical theory of selection,
recombination, and mutation.} Wiley Series in Mathematical and Computational
Biology. John Wiley \& Sons, Ltd., Chichester, 2000. xii+409 pp.

\bibitem{Eth1}  Etherington, I. M. H. Genetic algebras. Proc. Roy. Soc.
Edin. 59, 242-258, 1939.

\bibitem{Eth2}  Etherington, I. M. H. Special train algebras. Quart. J.
Math. (Oxford), 12, 1-8, 1941.

\bibitem{Eth3}  Etherington, I. M. H. Non-associative algebra and the
symbolism of genetics. Proc. Roy. Soc. Edin. B, 61, 24-42, 1941.

\bibitem{Ew}  Ewens, W. J. \emph{Mathematical population genetics.} \emph{I.
Theoretical introduction.} Second edition. Interdisciplinary Applied
Mathematics, 27. Springer-Verlag, New York, 2004.

\bibitem{FI}  Fran, F.; Irawati, I. The condition for a genetic algebra to
be a special train algebra. Journal of Multidisciplinary Engineering Science
and Technology. ISSN: 3159-0040, Vol. 2 Issue 6, 1496-1500, 2015.

\bibitem{GMR}  Ganikhodzhaev, R.; Mukhamedov, F.; Rozikov, U. Quadratic
stochastic operators and processes: results and open problems. Infin.
Dimens. Anal. Quantum. Probab. Relat. Top., 14 (2), 279-, 2011.

\bibitem{Gon1}  Gonshor, H. : Special train algebras arising in genetics.
Proc. Edinburgh Math. Soc. (2) 12, 41-53, 1960.

\bibitem{Gon2}  Gonshor, H. : Special train algebras arising in genetics,
II. Proc. Edinburgh Math. Soc. (2) 14, 333-338, 1965.

\bibitem{Hof}  Hofbauer, J.; Sigmund, K. \emph{Evolutionary games and
population dynamics.} Cambridge University Press, Cambridge, 1998.

\bibitem{Hol1}  Holgate, P. Genetic algebras associated with polyploidy.
Proc. Edinburgh Math. Soc. 15, 1-9, 1965.

\bibitem{Hol2}  Holgate, P. Population Algebras. J. R. Statist. Soc. B, 43,
No. 1, pp. 1-19, 1981.

\bibitem{Hol3}  Holgate, P. Some infinite-dimensional genetics algebras.
Alg\`{e}bres g\'{e}n\'{e}tiques (Montpellier, 1985), 35-45, Cahiers Math.
Montpellier, 38, Univ. Sci. Tech. Languedoc, Montpellier, 1989.

\bibitem{Karlin}  Karlin, S. Mathematical models, problems, and
controversies of evolutionary theory. Bull. Amer. Math. Soc. (N.S.) 10(2),
221-274, 1984.

\bibitem{Kes1}  Kesten, H. Quadratic Transformations: A Model for Population
Growth. I (and II). Advances in Applied Probability, Vol. 2, No. 1 (resp.
2), 1-82, (resp. 179-228) 1970.

\bibitem{Kes}  Kesten, H. Some nonlinear stochastic growth models. Bulletin
of the American Mathematical Society. Volume 77(4), 1971.

\bibitem{K1}  Kingman, J. F. C. \emph{Mathematics of genetic diversity.}
CBMS-NSF Regional Conference Series in Applied Mathematics, 34. Society for
Industrial and Applied Mathematics (SIAM), Philadelphia, Pa., 1980. vii+70
pp. ISBN: 0-89871-166-5.

\bibitem{K2}  Kingman, J. F. C. A matrix inequality. Quart. J. Math. Oxford
Ser. 12, 78-80, (1961).

\bibitem{Lyu}  Lyubich, Yu. I. \emph{Mathematical Structures in Population
Genetics.} Vol 22 of Biomathematics, Springer-Verlag, Berlin 1992.

\bibitem{MC1}  McCoy, N. H. On quasi-commutative matrices. Transactions of
the American Mathematical Society. Vol. 36(2), 327-340, 1934.

\bibitem{MC2}  McCoy, N. H. On the characteristic roots of matric
polynomials. Bull. Amer. Math. Soc. Vol. 42(8), 592-600, 1936.

\bibitem{Reed}  Reed, M. L. Algebraic structure of genetic inheritance. (New
Series) of the American Mathematical Society. Volume 34, Number 2, 107-130,
1997.

\bibitem{Weis}  Weissing, F. J., van Boven M. Selection and segregation
distortion in a sex-differentiated population. Theor Popul Biol., 60(4),
327-41, 2001.

\bibitem{Wo}  W\"{o}rz-Busekros, A. \emph{Algebras in Genetics.} Lecture
Notes in Biomathematics, Vol. 36. Springer, Berlin-Heidelberg-New York, 1980.

\bibitem{Wo2}  W\"{o}rz-Busekros, A. Relationship between genetic algebras
and semicommutative matrices. Linear Algebra and its Applications.
39,111-123, 1981.
\end{thebibliography}
\end{document}